\documentclass[journal,twoside]{IEEEtran}

\usepackage{cite}
\usepackage{amsmath,amssymb,amsfonts}
\usepackage{algorithmic}
\usepackage{graphicx}
\usepackage{textcomp}

\def\BibTeX{{\rm B\kern-.05em{\sc i\kern-.025em b}\kern-.08em
		T\kern-.1667em\lower.7ex\hbox{E}\kern-.125emX}}

\usepackage{amsmath,amssymb,amsfonts}
\usepackage{mathtools}

\usepackage{xcolor}
\usepackage{tikz,pgfplots}
\usepackage{bbding} %for loss/success symbol
\usepackage{ifthen} %for easier tikz-ing

\colorlet{darkGreen}{green!80!black}
\colorlet{lightblue}{blue!10!white}
\colorlet{darkOrange}{orange!90!black}
\colorlet{checkmarkGreen}{green!80!black}

\definecolor{istblue}{rgb/cmyk}{0.0,0.25490,0.56862745/1,0.55,0.0,0.43}%
\definecolor{istdarkgreen}{rgb/cmyk}{0.0627,0.5882,0.2824/0.89,0.0,0.52,0.41}%

\usetikzlibrary{arrows,arrows.meta,shapes,positioning,calc,decorations.pathreplacing,calligraphy}

\tikzstyle{smallCircle} = [draw, circle, fill=black,inner sep=0pt,outer sep=0.7pt,minimum size=3pt]
\tikzstyle{block} = [draw, rectangle,minimum height=3em, minimum width=3em]
\tikzstyle{sum} = [draw, circle]
\tikzstyle{box}=[rectangle, fill=gray!20, draw, minimum width=1.2cm, minimum height=0.5cm, align=center]
\tikzset{vertex/.style = {shape=circle,draw,minimum size=1.7em,inner sep=0}}
\tikzset{vertexFill/.style = {shape=circle,fill=gray!40!white,draw,minimum size=1.7em,inner sep=0}}
\tikzset{vertexFillMarked/.style = {shape=circle,fill=orange!50!white,draw,minimum size=1.7em,inner sep=0}}
\tikzset{edge/.style = {->,> = latex'}}
\tikzset{edgeMarked/.style = {->,> = latex',thick,darkOrange}}
\tikzset{lightning bolt to/.style={to path={
			let \p1=(\tikztostart), \p2=(\tikztotarget), \n1={veclen(\y2-\y1,\x2-\x1)} in
			(\p1) -- ($($(\p1)!0.6!(\p2)$)!\n1*.1!-90:(\p2)$) -- ($(\p1)!0.55!(\p2)$) --
			(\p2) -- ($($(\p1)!0.4!(\p2)$)!\n1*.1!90:(\p2)$) -- ($(\p1)!0.45!(\p2)$) -- 
			cycle (\p2)% Move to end point
}}}

\DeclarePairedDelimiter\abs{\lvert}{\rvert} %sometimes already defined
\newcommand{\ie}{i.e., }
\newcommand{\eg}{e.g., }
\newcommand{\cf}{cf.\ }

\newcommand{\anyWHRTC}[2]{\ensuremath{\genfrac{(}{)}{0pt}{}{#1}{#2}}}

\newcommand{\X}{\mathcal{X}}
\newcommand{\Xu}{\mathcal{X}_\mathrm{u}}
\newcommand{\Xo}{\mathcal{X}_0}
\newcommand{\B}{\Psi}

\newcommand*{\G}{\mathcal{G}}
\newcommand*{\U}{\mathcal{U}}
\newcommand*{\ua}{u^\mathrm{a}}
\newcommand*{\uc}{u^\mathrm{c}}
\newcommand*{\fc}{f{_\mathrm{c}}}

\newcommand*{\N}{\mathbb{N}}
\usepackage{bm}
\newcommand*{\sequ}[1]{\ensuremath{\bm{\mathbf{#1}}}}
\newcommand*{\R}{{\mathbb{R}}}
\newcommand*{\V}{\mathcal{V}}
\newcommand*{\E}{\mathcal{E}}

\newtheorem{proposition}{Proposition}

\newtheorem{remark}{Remark}
\newtheorem{theorem}{Theorem}
\newtheorem{corollary}{Corollary}
\newtheorem{definition}{Definition}

\newtheorem{problem}{Problem}
\newtheorem{casestudy}{Case Study}

\usepackage{siunitx}
\newcounter{examplecount}
\newtheorem{example}[examplecount]{Example}
\newcommand{\continuanceref}{}
\newtheorem{continuancex}{Example}

\newenvironment{continuance}[1] {\renewcommand\continuanceref{\ref{#1}}\continuancex} {\endcontinuancex}

\begin{document}
\title{
	Safety for Weakly-Hard Control Systems via Graph-Based Barrier Functions
}
\author{Marc Seidel, Mahathi Anand, Frank Allgöwer%
\thanks{Funded by the Deutsche Forschungsgemeinschaft (DFG, German Research Foundation) under Germany’s Excellence Strategy -- EXC2075 -- 390740016 and within grant AL 316/13-2 -- 285825138.}
\thanks{M.\ Seidel and F.\ Allg{\"o}wer are with the University of Stuttgart, Institute for Systems Theory and Automatic Control, Stuttgart, Germany, {\tt \small\{seidel,allgower\}@ist.uni-stuttgart.de}.}
\thanks{M.\ Anand is with the Chair of Robotics and Systems Intelligence, TU Munich, Germany, {\tt \small mahathi.anand@tum.de}.}
}

\maketitle

\begin{abstract}
	Despite significant advancement in technology, communication and computational failures are still prevalent in safety-critical engineering applications.
	Often, networked control systems experience packet dropouts, leading to open-loop behavior that significantly affects the behavior of the system.
	Similarly, in real-time control applications, control tasks frequently experience computational overruns and thus occasionally no new actuator command is issued.
	This article addresses the safety verification and controller synthesis problem for a class of control systems subject to weakly-hard constraints, i.e., a set of window-based constraints where the number of failures are bounded within a given time horizon.
	The results are based on a new notion of graph-based barrier functions that are specifically tailored to the considered system class, offering a set of constraints whose satisfaction leads to safety guarantees despite such failures.
	Subsequent reformulations of the safety constraints are proposed to alleviate conservatism and improve computational tractability, and the resulting trade-offs are discussed.
	Finally, several numerical case studies including linear and polynomial systems demonstrate the effectiveness of the proposed approach.
\end{abstract}


\section{Introduction} \label{sec:intro}
Real-time control systems are typically implemented on shared computational hardware, where computational tasks may be delayed due to deadline misses~\cite{Maggio2020}.
Fortunately, timely execution is not always required to achieve the desired control goal even for safety-critical tasks, as timing violations can be tolerated in many real systems \cite{Cervin2005,Akesson2020}.
Introduced in \cite{Bernat2001a}, the \emph{weakly-hard} (WH) model is an well-established abstraction to model tolerable deadline misses or failed executions.
Instead of requiring to meet every deadline, WH models allow for a bounded number of missed deadlines within a finite time interval.
This is typically formalized as so-called \emph{WH constraints} that constitute window-like temporal constraints bounding the pattern of deadline misses and hits \cite{Bernat2001a}.
These constraints also appear in the area of networked control systems, where packets sent over an unreliable communication channel are subject to occasional dropouts, \eg due to lossy communication links~\cite{Xiong2007, Blind2015, Linsenmayer2017}.
Such constraints can be validated on existing networks \cite{Ahrendts2018}, or guaranteed by design \cite{Broster2002,Linsenmayer2019a}.

However, a consequence of deadline misses or packet dropouts in a networked control system is in this case the unavailability of control inputs that need to be applied to the plant at all time.
As a result, the behavior of the plant may deviate from its desired response.
This is especially crucial in safety-critical applications, where pre-computed safe controllers may no longer be able to guarantee system safety in the wake of control unavailability.
This calls for an actuator design strategy that ensures intended behavior despite control losses. In this regard, two simple actuator strategies emerge: \emph{zero} or \emph{hold}~\cite{Schenato2009}.
While, the zero strategy sets the actuator to zero in the event of a deadline miss or a packet dropout, the hold strategy utilizes the last available input to compensate for the loss.
While WH constraints prescribe the allowable number of deadline misses in a networked control system or real-time control system (also called \emph{WH control systems}), the actuator strategy dictates how the system handles the control input losses when they occur.
Together, they yield a system model that is amenable to formal analysis and controller synthesis.
In this paper, we propose a formal analysis and controller synthesis framework for the safety of WH control systems.

\textbf{Related Work.}
WH systems and the notion of WH constraints were first introduced in \cite{Bernat2001a} as class of real-time systems.
Since then, graph-based representations for WH constraints have been widely adopted, particularly to simplify the mathematical description, enabling formal analysis of WH control systems~\cite{Horssen2016,Hamdaoui1995,Vreman2022c,Linsenmayer2017,Huang2019a}.
Specifically, the paths along the graph describe deadline misses or execution failures, providing predictions for future scenarios.
These are then accounted for during system design, e.g., in control theory, where WH constraints model input losses~\cite{Linsenmayer2017}.
Literature in this context is primarily concerned with the study of stability properties, where graph-based representations are combined with joint spectral radius~\cite{Maggio2020,Vreman2022b} or Lyapunov-based analysis~\cite{Blind2015,Linsenmayer2017,Linsenmayer2021a}, or with controller design subject to performance objectives~\cite{Seidel2024b,Vreman2021,Pazzaglia2018,Lang2024}.

Contrary to stability analysis, notions of safety for WH control systems remain largely unexplored and existing results are either tailored to specific system classes or use restricted definitions of safety.
For example, \cite{Hobbs2022,Xu2023,Banerjee2025} consider a safety property defined as the maximum deviation from a nominal trajectory, while~\cite{Huang2019a} define safety as the trajectories staying within a predefined ball around the system's equilibrium.
A more general notion of safety is considered in~\cite{Huang2020} to obtain a set of initial conditions from which the trajectories remain bounded to a safe set.
However, these works only consider safety verification and are either limited to linear systems or rely on state space discretizations, making them computationally intractable for general nonlinear systems.
An effective alternative is to utilize barrier functions~\cite{Prajna2004}.
Building upon Lyapunov-based techniques, they provide sufficient conditions for ensuring that the trajectories starting at a specified initial set do not reach unsafe regions.
Recently, barrier functions have been utilized for both safety verification and controller synthesis of nonlinear systems~\cite{Prajna2006,Ames2014} as well as switched and hybrid systems~\cite{Prajna2004, cbf_switch1, asymp_stab_safe, state-time-safety}.
Recently, more general graph-based safety conditions using barrier functions were proposed in~\cite{Anand2024a} for switched systems.
Here, barrier functions are defined for each node of the graph, such that graph edges correspond to a set of algebraic conditions imposed on the barrier functions, whose satisfaction then leads to the verification of safety constraints.

\textbf{Contributions.}
In this article, we address the existing gaps in the literature by proposing for the first time a barrier function-based approach for the safety verification and controller synthesis for WH control systems, i.e., control systems experiencing dropouts and deadline misses subject to WH constraints.
In particular, inspired by~\cite{Anand2024a}, we propose a new notion of graph-based barrier functions (GBFs) that are tailored to the underlying WH constraints.
To do so, we first represent WH constraints using appropriate graphs, where the edges provide information on the possibility of losses in the future.
This information is incorporated into safety conditions by the GBF---a collection of barrier functions, one for each node of the graph---that is required to satisfy a number of algebraic conditions that depend on the corresponding edge information.
Note that the obtained results are general, non-conservative, can be applied to WH control systems operating under both zero and hold actuator strategies, and are applicable for controller synthesis as well as verification.
However, the algebraic conditions presented by GBFs require recursive computation of system dynamics and satisfaction logical implication conditions, which may not always be computationally tractable for complex systems.
Therefore, we provide alternative reformulations that alleviate the aforementioned issues at the cost of conservatism.
Finally, we summarize the numerical computation of graph-based barrier functions for linear and polynomial WH control systems, and validate our results through case studies, and qualitatively compare the conservatism presented by the different formulations through simulation.

The approach presented here is also closely related to the one in~\cite{Anand2021}, where flexible safety conditions via logical implications and $k$-induction are proposed.
However, our work differs from it significantly.
First, \cite{Anand2021} is limited to safety verification of nonlinear dynamical systems, while we consider verification and controller synthesis for WH control systems.
Secondly, our work incorporates barrier functions with graphs that result from system specifications, i.e., WH constraints, whereas no such considerations are made in~\cite{Anand2021}.
Further connections between these works, including differences, are highlighted where necessary in the paper.

The remainder of this article is structured as follows.
We provide basic notation, definitions, and the problem statement in Section~\ref{sec:setup}.
Section~\ref{sec:prelim} introduces some preliminary results on WH constraints and barrier functions.
The main theoretical results concerning the formulation of GBFs for safety verification and controller synthesis are presented in Section~\ref{sec:safety}.
In Section~\ref{sec:computation}, we discuss the computation of GBFs, and we subsequently provide numerical case studies to demonstrate the effectiveness of our results in Section~\ref{sec:numerics}.
Section~\ref{sec:conclusion} concludes the article.

\section{Problem Formulation} \label{sec:setup}

\subsection{Notation}
The set of real and non-negative integers are denoted by $\R$ and $\N$, respectively.
$\R^n$ denotes a real space of dimension $n$ and $\R^{m \times n}$ denotes a real space of dimension $m \times n$.
Matrices are denoted in uppercase, e.g., $A \in \R^{m \times n}$, and vectors in lowercase $x \in \R^n$.
The transpose of a vector or matrix is $(\,)^\top$.
For two sets $\mathcal{A}$ and $\mathcal{B}$, a function $f\colon \mathcal{A} \rightarrow \mathcal{B}$ is a mapping from $\mathcal{A}$ to $\mathcal{B}$, and $\mathrm{id}_\mathcal{A}$ is the identity function on the set $\mathcal{A}$.
Given three sets $\mathcal{A},\mathcal{B},\mathcal{C}$ and functions $f\colon \mathcal{A} \rightarrow \mathcal{B}$ and $g\colon \mathcal{B} \rightarrow \mathcal{C}$, we denote the composition of functions $f$ and $g$ by $g \circ f\colon \mathcal{A} \rightarrow \mathcal{C}$.
For a function $f\colon \mathcal{A} \rightarrow \mathcal{A}$ and $n \in \mathbb{N}$, $f^i$ denotes the $i^{\mathrm{th}}$ iterate of the function $f$, defined as $f^{n}=\mathrm{id}_\mathcal{A}$ when $n=0$ and $f^{n}=f^{n-1} \circ f$ when $n \neq 0$.
Finally, given a (in)finite set $\mathcal{A}$, a sequence (denoted by bold symbols) $\sequ{w} = w_1 w_2 \ldots$ is an infinite concatenation of letters, \ie $w_i \in \mathcal{A}$, for all $i \geq 0$.
We also use $\sequ{w}(i) = w_i$ to denote the $i^{\mathrm{th}}$ position of the sequence $\sequ{w}$.
Moreover, a subsequence of $\sequ{w}$, denoted by $\sequ{w}'$ is given by  $\sequ{w}' = w_i \ldots w_{k}$, for any $ 1 \leq i \leq k$.

\subsection{Weakly-Hard Control Systems}
\label{sec:setup-WHCS}

We consider discrete-time, nonlinear control systems of the form
\begin{align}
	\sequ x(t+1) = f(\sequ x(t), \sequ{u}^\mathrm{a}(t)), \label{eq:sys-nonl}
\end{align}
where $\sequ x(t) \in \X \subseteq \mathbb{R}^n$ and $\sequ{u}^\mathrm{a}(t) \in \U \subseteq \mathbb{R}^{n_u}$ are the state and applied (actuator) inputs at time $t \in \mathbb{N}$, with $\X$ and $\U$ being the corresponding state and  input sets, respectively.
The system is controlled by a nonlinear, static state-feedback controller that computes the control input at time $t$ as
\begin{align}
	\sequ{u}^\mathrm{c}(t) = g(\sequ{x}(t)). \label{eq:controller}
\end{align}
Unfortunately, the desired control input may not be available at all times due to transmission losses or unavailability of shared resources that result in a deadline miss.
This is particularly relevant in settings like networked systems where a control signal is sent via faulty communication networks, or in real-time control applications where one utilizes shared computational resources for improved efficiency.
We describe the losses by a loss sequence $\sequ{\mu}= \sequ{\mu}(1)\sequ{\mu}(2)\ldots$, such that at $t\in \N_0$, we have $\sequ{\mu}(t) = 1$ if a new value of $\sequ{u}^\mathrm{c}(t)$ at time $t$ is available and $\sequ{\mu}(t) = 0$ otherwise.
For simplicity, we say that there is a \emph{success} at time $t$ when $\sequ{\mu}(t) = 1$, and a \emph{loss} otherwise.
Moreover, without any loss of generality, we assume that $	\sequ{\mu}(0)=1$.

There are two common approaches to compute the applied actuator input in the presence of losses~\cite{Schenato2009}.
The first is to use the \emph{zero} strategy, where zero input is applied in the case of a loss. This can be formally described as
\begin{align}\label{eq:actuator-input-zero}
	\raisetag{3.5em}
	\sequ{u}^\mathrm{a}_z(t) = \begin{cases}
		\sequ{u}^\mathrm{c}(t) \quad & \text{if $\sequ{\mu}(t) = 1$,}\\
		0 \quad &\text{if $\sequ{\mu}(t) = 0$.} \\
	\end{cases}
\end{align}
The second is the \emph{hold} strategy, where the last successful control input is held until another success occurs.
Here, we define the applied input recursively as
\begin{align}\label{eq:actuator-input-hold}
	\sequ{u}^\mathrm{a}_h(t) = \begin{cases}
		\sequ{u}^\mathrm{c}(t) \quad & \text{if $\sequ{\mu}(t) = 1$,}\\
		\sequ{u}^\mathrm{a}_h(t-1) \quad &\text{if $\sequ{\mu}(t) = 0$.} \\
	\end{cases}
\end{align}
Finally, given an initial state $\sequ x(0)=x_0$, a state-feedback controller $g$, a loss sequence $\sequ{\mu}$, and a pre-determined control strategy (\ie zero or hold), one obtains the state sequence as $\sequ{x}_{g}^{x_0} = x_0 \sequ x(1) \sequ x(2) \ldots$ by utilizing the corresponding control input sequence $\sequ{u}^\mathrm{c} = g(x_0)g(\sequ x(1))g(\sequ x(2))\ldots$ and the applied input sequence $\sequ{u}_q^\mathrm{a} = \sequ{u}^\mathrm{a}_q(0)\sequ{u}^\mathrm{a}_q(1)\sequ{u}^\mathrm{a}_q(2)\ldots$ computed according to~\eqref{eq:actuator-input-zero} (when $q = z$) or~\eqref{eq:actuator-input-hold} (when $q=h$), respectively.
An overview of the control system in the presence of losses can be found in Figure~\ref{fig:blockDiagram-setup}.
The system consists of an open-loop system~\eqref{eq:sys-nonl} and a state feedback controller~\eqref{eq:controller}.
Moreover, $\sequ{\mu}(t)$ at time $t \in \N$ can be viewed as a switch that prevents the control input $\sequ{u}^\mathrm{c}(t)$ from being applied to the actuator whenever $\sequ \mu(t)=0$ for some $t\in \N$.

\begin{figure}
	\centering
	\begin{tikzpicture}[auto,>=latex,thick]
	\node [block,align=center] (sys) {System \\ $\sequ{x}(t+1) = f(\sequ{x}(t),\sequ{u}^\mathrm{a}_{q}(t))$};
	\node [left=of sys,shift={(-0.7,0)},outer sep=0.3em](drop) {};
	\node [block, below=of sys, align=center, yshift=1.4em] (K) {Controller \\ $\sequ{u}^\mathrm{c}(t) = g(\sequ{x}(t))$};
	\draw (K.west) -| node[above, pos=0.25] {$\sequ{u}^\mathrm{c}(t)$} ($(drop.west)+(-0.5,0.0)$) -- (drop.west);
	\draw (drop.west) -- node[above, pos=0.5] {$\sequ{\mu}(t)$} (drop.north east);
	\draw [-latex] (drop.east) --  node[below, pos=0.5] {$\sequ{u}^\mathrm{a}_{q}(t)$} (sys.west);
	\draw [-latex] (sys.east) -- ($(sys.east)+(0.5,0.0)$) |- node[right, pos=0.25] {$\sequ{x}(t)$} (K.east);
\end{tikzpicture}
	\caption{The control system with losses in the feedback loop.}
	\label{fig:blockDiagram-setup}
\end{figure}

We make a few more notational modifications to simplify the presentation in the article.
In particular, we rewrite the closed-loop system dynamics under the zero strategy as
\begin{align} \label{eq:sys-zero}
	\sequ x(t+1) = \sequ \mu(t) \fc(\sequ x(t)) + (1-\sequ \mu(t)) f_{\mathrm{o}_z}(\sequ x(t)),
\end{align}
where $\fc(x) = f(x,g(x))$ and $f_{\mathrm{o}_z}(x) = f(x, 0)$.
Control systems using the hold strategy may be formulated in two different ways.
The first is a memory-dependent system representation given by
\begin{equation} \label{eq:sys-hold1}
	\sequ x(t+1) = \sequ \mu(t) \fc(x) + (1- \sequ \mu(t)) f_{\mathrm{o}_h}(\sequ x(t), \sequ x(t-t_h))), 
\end{equation}
where $\fc$ is as defined above and $f_{\mathrm{o}_h}(x,y) = f(x, g(y))$ and $t_h \coloneq \{\, \min t_i \in \mathbb{N} \mid \sequ{\mu}(t-t_i) = 1 \,\}$.
However, with a slight abuse of notation, we remove the second argument from $f_{\mathrm{o}_h}$ and write $f_{\mathrm{o}_h}(x) = f(x,g(x_h))$, where $x_h$ is the last held state at time $t-t_h$ that is implicitly dependent on $x$.
Then, one can use $\fc$ and $f_{\mathrm{o}_q}, \, q \in \{z,h\}$, to denote the closed-loop and open-loop dynamics of control systems with zero and hold strategies, respectively.
The second formulation for control systems with the hold strategy utilizes an augmented state $\tilde x = \begin{bmatrix} x & \ua_h \end{bmatrix}^\top$ to obtain
\begin{equation} \label{eq:sys-hold2}
	\tilde{\sequ x}(t+1) = \sequ \mu(t) \tilde{f}_{\mathrm{c}_h}(\tilde{\sequ x}(t)) + (1 - \sequ \mu(t)) \tilde{f}_{\mathrm{o}_h}(\tilde{\sequ x}(t)),
\end{equation}
where $\tilde{\fc}(\tilde x) = \begin{bmatrix}
	f(x, g(x)) \\
	g(x) \end{bmatrix}$ and  $\tilde{f}_{\mathrm{o}_h}(\tilde{x}) = \begin{bmatrix}
	f(x, {u}^\mathrm{a}_h) \\
	{u}^\mathrm{a}_h
	\end{bmatrix}$.
The usage of the appropriate formulation of the system with the hold strategy will be made clear with context.

In many practical scenarios, losses in control inputs are not completely random, but can instead be modeled via a set of deterministic constraints.
In this article, we consider WH constraints, adapted from~\cite{Bernat2001a}, which constrain the permissible number of losses that can occur in a specified time window.

\begin{definition}[WH Constraint] \label{def:WHC}
	A loss sequence $\sequ{\mu}$ satisfies the WH constraint \anyWHRTC{r}{s} (``meets $r$ in $s$'', $r$, $s \in \mathbb{N}$, $r \leq s$), if in any window of $s$ consecutive control attempts, there are at least $r$ of them successful (in any order).
\end{definition}
The following example seeks to illustrate the class of WH constraints considered in the article.

\begin{example}\label{ex:runningEx}
	The loss sequence $\sequ{\mu} = 1 \, 0 \, 0 \, 1 \, 1 \, 0  \ldots$ satisfies the WH constraint \anyWHRTC{2}{4}, since every window of four consecutive time instants has at least two successes.
\end{example}

We now formally define \emph{weakly-hard control systems}, which are the main subject of this work.

\begin{definition}[WH Control System] \label{def:WHCS}
	A WH control system is the conjunction of system~\eqref{eq:sys-nonl} and controller~\eqref{eq:controller} (see Figure~\ref{fig:blockDiagram-setup}) under loss sequences satisfying a WH constraint \anyWHRTC{r}{s}.
\end{definition}

\begin{remark}
	Other descriptions of WH constraints, such as by constraining the number of (consecutive) losses or successes are possible, and our work naturally extends to these cases.
	For more details, see~\cite{Bernat2001a}.
	Note that the methodology presented in this work is generally applicable to the other variations of Definition~\ref{def:WHC}.
	For ease of exposition, we omit the study of these cases.
\end{remark}

\begin{remark}
	Note that Definition~\ref{def:WHCS} is independent of the actuator strategy (zero or hold) considered.
	In the following sections, we shall explicitly indicate the employed strategy when necessary.
	Otherwise, either strategy may be applied.
\end{remark}

In the following, we introduce the safety verification and synthesis problem for WH control systems, which is the main topic of the article.

\subsection{Safety} \label{sec:setup-safetyDef}

A WH control system is said to be safe if it can effectively avoid any undesirable configurations irrespective of the losses occurring in the system.
A formal definition follows.

\begin{definition}[Safety of WH Control Systems] \label{def:safety}
	Consider a WH control system as in Definition~\ref{def:WHCS}.
	Let $\Xo \subseteq \X$ be the initial set and $\Xu \subseteq \X$ be the unsafe set of the system such that $\Xo \cap \Xu = \emptyset$.
	Then, a WH control system is said to be safe w.r.t.\ $\Xo$ and $\Xu$ under a controller $g$ if for any $x_0$ and any loss sequence $\sequ{\mu}$ that satisfies a WH constraint as in Definition~\ref{def:WHC}, we have that $\sequ{x}_{g}^{x_0}(t) \notin \Xu, \forall t \in \N$.
\end{definition}

In this article, we are concerned with safety \emph{verification} and \emph{synthesis} problems for WH control systems employing both zero and hold strategies.
Safety verification entails checking whether a WH control system is safe under a given controller $g$.
The synthesis problem asks to compute a suitable controller $g$ that renders the WH control system safe.
Formally, this is given as follows.

\begin{problem}[Safety Verification] \label{prob:safetyVerification}
	Given a WH control system under a controller $g$ with $g(x) \in \U \enspace \forall x \in \X$, a WH constraint \anyWHRTC{r}{s}, and sets $\Xo$, $\Xu$, verify whether the system is safe w.r.t.\ $\Xo$ and $\Xu$, \ie show that $\sequ{x}_{g}^{x_0}(t) \notin \Xu, \forall t \in \N$, for any $x_0 \in \Xo$.
\end{problem}

\begin{problem}[Safe Controller Synthesis] \label{prob:safetySynthesis}
	Given a WH control system, a WH constraint $\anyWHRTC{r}{s}$, and sets $\Xo$, $\Xu$, compute a suitable controller $g(x) \in \U$ s.\ t.\ the system is safe w.r.t.\ $\Xo$ and $\Xu$, \ie find $g$ such that $\sequ{x}_{g}^{x_0}(t) \notin \Xu, \forall t \in \N$, for any $x_0 \in \Xo$.
\end{problem}

\section{Preliminaries} \label{sec:prelim}
Some essential preliminary concepts are introduced below.
First, we describe the systematic representation of WH constraints via graphs.
Then, we introduce the notion of barrier functions, which together with the graphs, allow us to determine safety conditions efficiently.

\subsection{Weakly-hard graphs} \label{sec:prelim-WHgraph}
In this section, we integrate the possible loss sequences into the system description via graph representations of WH constraints. Before introducing the graph, without loss of generality, we first divide a loss sequence satisfying a WH constraint \anyWHRTC{r}{s} into meaningful subsequences such that each subsequence starts with a $1$ (success) and is followed by a number of consecutive $0$'s (losses).
In particular, a loss sequence $\sequ{\mu}$ can be written as $\sequ{\mu} = \sequ{\mu}'_1 \sequ{\mu}'_2 \ldots$, where the length $k_i$ of subsequence $\sequ{\mu}'_i$, $i \in \N$, is at most $s-r + 1$. Moreover, for each subsequence, $\sequ{\mu}_i'(0) = 1$ and $\sequ{\mu}_i'(j) = 0$, for $j \in \{1,\ldots,k_i\}$.
To concisely refer to these subsequences, we assign the label $l_i = k_i-1$ to each $\sequ{\mu}_i'$; this is nothing but the number of consecutive losses $l_i$, \ie the number of $0$'s in $\sequ{\mu}_i'$.
As a result, each label $l_i$ denotes a unique corresponding subsequence with $l_i$ consecutive losses, \eg $l_i=2$ represents $1 \, 0 \, 0$.
Note that the number of possible labels for a given WH constraint $\anyWHRTC{r}{s}$ is $s-r+1$, which is the maximum number of consecutive losses allowed for \anyWHRTC{r}{s} plus label $0$ for no losses.
Therefore, one has $l_i \in \Sigma$ with $\Sigma = \{0,\ldots,s-r\}$ being the finite alphabet of all possible labels.

Now, consider a directed graph $\G = (\mathcal{V},\mathcal{E})$ with $\mathcal{V}$ as the set of nodes and $\mathcal{E}$ as the set of edges.
An edge $(v,l,v') \in \mathcal{E}$ starts from node $v$ and ends in node $v'$ with label $l \in \Sigma$, i.e., it represents the number of consecutive losses encountered since the last successful attempt, while the nodes implicitly encode the relevant loss history.
A path in the graph $\G$ is a sequence of nodes and edges $\pi = v^0 l^0 v^1 l^1 v^2 \ldots$ such that $(v^i, l^i, v^{i+1}) \in \mathcal{E}, \forall i \in \N$.
A sequence of labels $l^0 l^1 \ldots$ is said to be accepted by the graph if there exists a corresponding path $\pi$ in $\G$.
The language of $\G$ is the set of all sequences that are accepted by $\G$.
The graph $\G$, called a \emph{WH graph}, corresponds to a WH constraint \anyWHRTC{r}{s} if the language of $\G$ is equivalent to the set of all loss sequences that satisfy \anyWHRTC{r}{s}.
One can see that each edge of $\G$ captures one of the possible subsequences $\sequ{\mu}'$ of the constraint \anyWHRTC{r}{s}, represented by the corresponding label $l \in \Sigma$.
As a consequence, all possible loss sequences can be encoded by moving along the edges of $\G$, \ie via the paths of $\G$.
Similarly, all sequences accepted by $\G$ satisfy the WH constraint \anyWHRTC{r}{s}.

\begin{remark}
	The presented definition of WH graphs slightly differs from some earlier works \cite{Vreman2022c, Seidel2024b}. However, it corresponds to the notion of a \emph{lifted graph} that provides a minimal representation to a WH constraint, see \cite{Seidel2024b} for a comparison and discussion.
	Algorithms for generating such graphs for a given WH constraint can be found in \cite{Linsenmayer2021a,Vreman2022c}.
\end{remark}

\begin{continuance}{ex:runningEx}[continued]
	For the WH constraint \anyWHRTC{2}{4}, one has $\Sigma = \{0,1,2\}$.
	The loss sequence $\sequ{\mu} = 1 \, 0 \, 0 \, 1 \, 1 \, 0  \ldots$ is divided into three subsequences:
	$\sequ{\mu} = \sequ{\mu}_1' \sequ{\mu}_2' \sequ{\mu}_3'$ with $\sequ{\mu}_1' = 1 \, 0 \, 0$, $\sequ{\mu}_2' = 1$, $\sequ{\mu}_3' = 1 \, 0$, and their respective labels $l_1=2$, $l_2=0$, $l_3=1$.
	Moreover, the graph for the constraint \anyWHRTC{2}{4} is depicted in Figure~\ref{fig:WHgraph-example} and $\sequ{\mu}$ corresponds to the path $v_1 2 v_3 0 v_1 1 v_2 \ldots$, transitioning via the edges $(v_1,2,v_3)$, $(v_3,0,v_1)$, and $(v_1,1,v_2)$.%
	\begin{figure}
		\centering
		\begin{tikzpicture}
%\tikzset{vertex/.style = {shape=circle,draw,minimum size=1.7em,inner sep=0}}
%\tikzset{vertexFill/.style = {shape=circle,fill=gray!50!white,draw,minimum size=1.6em,inner sep=0}}
%\tikzset{edge/.style = {->,> = latex'}}

%\begin{scope}[shift={(0,0)}]
%	%vertices
%	\node[vertex] (v1) at (0,0) {$v_1$};
%	\node[vertex] (v2) at (-0.75,-1) {$v_2$};
%	\node[vertex] (v3) at (0.75,-1) {$v_3$};
%	\node[vertex] (v4) at (-1.1,0.1) {$v_4$};
%	\node[vertex] (v5) at (1.1,0.7) {$v_5$};
%	\node[vertex] (v6) at (1.75,-0.3) {$v_6$};
%	\node[vertex] (v7) at (-2,-1) {$v_7$};
%	
%	%edges
%	\draw[edge] (v1) to[loop above] node[above, pos=0.5] {\footnotesize$1$} (v1);
%	\draw[edge] (v1) to[bend right] node[above, pos=0.5] {\footnotesize$0$} (v4);
%	\draw[edge] (v4) to[bend right] node[right, pos=0.3] {\footnotesize$1$} (v2);
%	\draw[edge] (v2) to[bend right] node[left, pos=0.7] {\footnotesize$1$} (v1);
%	\draw[edge] (v2) to[bend right] node[above, pos=0.6] {\footnotesize$0$} (v7);
%	\draw[edge] (v7) to[bend right] node[below, pos=0.5] {\footnotesize$1$} (v2);
%	\draw[edge] (v1) to[bend left] node[above, pos=0.5] {\footnotesize$0$} (v5);
%	\draw[edge] (v5) to[bend left] node[above, pos=0.5] {\footnotesize$0$} (v6);
%	\draw[edge] (v6) to[bend left] node[below, pos=0.3] {\footnotesize$1$} (v3);
%	\draw[edge] (v3) to[bend left] node[below, pos=0.5] {\footnotesize$1$} (v1);
%	
%	%constraint
%	%\node (WHRT23) at ($(0,0)+(0,-1.9)$) {$\mathcal{G}$ for $\lambda = \anyWHRTC{2}{4}$};
%\end{scope}

\begin{scope}[shift={(0,0)}]
	%vertices
	\node[vertexFill] (v1) at (0,0) {$v_1$};
	\node[vertexFill] (v2) at (-1.7,0) {$v_2$};
	\node[vertexFill] (v3) at (1.7,0) {$v_3$};
	
	%edges
	\draw[edge] (v1) to[loop above] node[above, pos=0.5] {\footnotesize$0$} (v1);
	\draw[edge] (v1) to[bend right] node[above, pos=0.5] {\footnotesize$1$} (v2);
	\draw[edge] (v2) to[bend right] node[below, pos=0.5] {\footnotesize$0$} (v1);
	\draw[edge] (v2) to[loop above] node[above, pos=0.5] {\footnotesize$1$} (v2);
	\draw[edge] (v1) to[bend left] node[above, pos=0.5] {\footnotesize$2$} (v3);
	\draw[edge] (v3) to[bend left] node[below, pos=0.5] {\footnotesize$0$} (v1);
	
	%constraint
	%\node (WHRT23) at ($(0,0)+(0,-1.9)$) {$\tilde{\mathcal{G}}$ for $\lambda = \anyWHRTC{2}{4}$};
	
	\node[align=left] (legend) at (-4,0.2) {\footnotesize labels $l$ representing\\[-0.5ex] \footnotesize subsequences $\bm \mu'$:\\ \footnotesize $l = 0\colon \sequ{\mu}' = 1$ \\[-0.2ex] \footnotesize $l = 1 \colon \sequ{\mu}' = 1 \, 0$ \\[-0.2ex] \footnotesize $l = 2\colon \sequ{\mu}' = 1 \, 0 \, 0$};
\end{scope}

\end{tikzpicture}
		\caption{WH graph $\G$ for the WH constraint \anyWHRTC{2}{4}, and its labels representing the respective subsequences.}
		\label{fig:WHgraph-example}
	\end{figure}
\end{continuance}

\subsection{Barrier functions as safety certificates} \label{sec:prelim-barrierFunctions}
Barrier functions \cite{Prajna2004} are real-valued functions defined over the state set $\X$, such that their zero-level sets act as a barrier separating the safe and the unsafe regions of the state set.
If one can show that the state sequences of the system never cross this level set, safety of the system can be concluded.

For the purpose of introduction, let us first consider the loss-free control system, that is \eqref{eq:sys-nonl} with controller \eqref{eq:controller} and $\sequ{u}^\mathrm{a}(t) = \sequ{u}^\mathrm{c}(t)$ $\forall t \in \N$, resulting in the dynamics $\sequ x(t+1) = \fc(\sequ x(t))$.
Note that this is equivalent to a WH control system with $\sequ{\mu}(t) = 1$ $\forall t \in \mathbb{N}$, \ie under the trivial constraint \anyWHRTC{1}{1}.
Then, barrier functions can be defined as follows.

\begin{definition} \label{def:basic-BF}
	For a WH control system under the WH constraint $\anyWHRTC{1}{1}$, a function $\B \colon \X \rightarrow \mathbb{R}$ is called \emph{barrier function} w.r.t.\ an initial set $\Xo$ and unsafe set $\Xu$ if the following conditions hold:%
	\begin{subequations}\label{eq:safety-lossFree}
		\begin{alignat}{2}
			\B(x) \leq{}& 0   \quad &&\forall x \in \Xo \label{eq:safety-lossFree-initialSet}\\
			\B(x) >{}& 0   \quad &&\forall x \in \Xu \label{eq:safety-lossFree-unsafeSet}\\
			\B(x) \leq{}& 0 ~ \Rightarrow ~ \B( \fc(x) ) \leq 0 \quad &&\forall x \in \X \label{eq:safety-lossFree-implication}
		\end{alignat}
	\end{subequations}
\end{definition}

\begin{proposition} \label{thm:safety-lossFree}
	The WH control system under the WH constraint $\anyWHRTC{1}{1}$ is safe with respect to $\Xo$ and $\Xu$ if there exists a barrier function according to Definition~\ref{def:basic-BF}.
\end{proposition}

The intuition behind \eqref{eq:safety-lossFree} is as follows.
Conditions~\eqref{eq:safety-lossFree-initialSet} and~\eqref{eq:safety-lossFree-unsafeSet} establish the level set $\B(x) = 0$ to act as a barrier that separates the safe states from the unsafe states, and condition~\eqref{eq:safety-lossFree-implication} ensures that the the system always stays within the zero-level set, thus guaranteeing safety of the system.
The above result is for verification of a system under a given controller $g(x)$.
However, if the controller is unknown and must be designed, one seeks to find $g$ and $\B$ such that~\eqref{eq:safety-lossFree-implication} holds.

\begin{remark}
	Note that similar safety conditions were first presented in~\cite{Prajna2004} in the context of hybrid systems.
	However, those conditions are slightly more restrictive than the ones in~\eqref{eq:safety-lossFree} as they ask for the value of the barrier function to decrease at every time step.
	In contrast, our conditions allow the value of the barrier function to increase over each time step as long as it is within the zero-sublevel set.
	These are equivalent to the safety conditions presented in~\cite{Anand2021}.
\end{remark}

Unfortunately, conditions~\eqref{eq:safety-lossFree-initialSet}--\eqref{eq:safety-lossFree-implication} are no longer valid in the presence of losses.
Particularly, loss events render the controller $g(x)$ inaccessible and consequently,~\eqref{eq:safety-lossFree-implication} may not hold at every time step, potentially leading to violation of safety.
In the following section, we seek to alleviate this issue by presenting a novel graph-based barrier function approach for the safety verification (Problem~\ref{prob:safetyVerification}) and synthesis (Problem~\ref{prob:safetySynthesis}) for general WH control systems by utilizing graph representations of WH constraints.

\section{Safety Verification and Synthesis} \label{sec:safety}

In this section, we generalize~\eqref{eq:safety-lossFree} to WH control systems under arbitrary WH constraints by incorporating safety conditions into WH graphs.
In particular, we introduce the notion of \emph{graph-based barrier functions} that define a set of safety conditions along the edges of the WH graph.
By doing so, one ensures safety for all the possible loss sequences that are encoded within the graph.
We first deal with the safety verification problem, followed by controller synthesis.

\subsection{Safety Verification of WH Control Systems} \label{sec:safety-verification}
Consider a WH control system according to Definition~\ref{def:WHCS}.
Since the losses in the system are quasi-deterministic, i.e., they are bounded by a permissible number of losses determined by the WH constraint, it is possible to adapt conditions~\eqref{eq:safety-lossFree} to ensure that the system is steered sufficiently far away from the unsafe states whenever there is a success and a controller is correctly applied to the actuator.
This serves as a buffer in preparation for the possible following losses, potentially driving the system towards the unsafe set.

More formally, consider a WH control system operating under a constraint $\anyWHRTC{r}{s}$, and a time $t \in \N$ such that $\sequ{\mu}(t)~=~1$.
Due to the WH constraint, this success may be followed by $l \in \{0,\ldots,s-r\}$ losses.
Suppose that at time $t$, the value of the barrier function is sufficiently negative.
Then, at each time instant $t+m$ with $\sequ{\mu}(t+m) = 0$, $m \leq l$, the barrier function may be permitted to take up larger values as long as the zero-level set of the barrier function is not crossed.
Recall that that each edge in the WH graph $\G = (\mathcal{V},\mathcal{E})$ of the constraint $\anyWHRTC{r}{s}$ corresponds to one success followed by $l$ losses.
Therefore, by imposing the above condition to each edge of $\G$, one ensures the inductive application of the prescribed barrier function behavior along all the paths of the graph.
This establishes safety of the WH control system robustly for all the possible loss sequences that the system may encounter.

Although the requirement described above is sufficient for the safety of WH control systems, it does not fully utilize the information available from the graph.
In particular, the WH graph implicitly contains information on the number of losses possible in the future, \ie each edge $(v,l,v') \in \E$ of the graph $\G$ encodes the possibility of exactly $l$ number of losses, which restricts the number of future consecutive losses and is bounded by $s-r-l$.
Thus, the edge labels on the edges outgoing from $v'$ are also bounded.
Accordingly, the barrier function values may be adjusted over time depending on the possibility of future successes or losses.
In other words, if it is known that the next control attempt is to be successful, one can allow the system's state to be closer to the unsafe set $\Xu$ by assigning a larger barrier function to the node.
An illustration of how WH graphs encode the information about subsequent losses is given below.

\begin{continuance}{ex:runningEx}[continued]
	At node $v_3$ of $\G$ in Figure~\ref{fig:WHgraph-example}, it is known that there have already been two consecutive losses in the past, and as a result, two successes are guaranteed in the next two time instants, as encoded by the edges $(v_3,0,v_1)$, and $(v_1,0,v_1)$, $(v_1,1,v_2)$ or $(v_1,2,v_3)$, respectively.
\end{continuance}

In order to encode the aforementioned behavior, we utilize so-called \emph{graph-based barrier functions}, wherein each node $v \in \mathcal{V}$ of the WH graph $\G$ is assigned a barrier function $\B_{v}$.
An edge traversal $(v,l,v') \in \E$ in $\G$ is triggered with a success, and consequently, a switch is made from $\B_{v}$ to $\B_{v'}$.
The values taken by $\B_{v'}$ are adjusted according to $l$ with respect to $\B_{v}$, thereby ensuring adaptability of safety conditions to the possible number of future losses.
We now formally define graph-based barrier functions.

\begin{definition}[Graph-based Barrier Function] \label{def:graph-based-BF}
	Consider a WH control system with the WH graph $\G = (\mathcal{V},\mathcal{E})$ and $g(x) \in \U \enspace \forall x \in \X$. %
	Suppose that the loss sequences satisfy the WH constraint \anyWHRTC{r}{s} with the corresponding WH graph $\G = (\mathcal{V},\mathcal{E})$.
	A set of $n_\mathcal{V}$ functions $\B_v(x)\colon \X \rightarrow \mathbb{R}$ is called a \emph{graph-based barrier function} (GBF) for $\G$ w.r.t.\ an initial set $\Xo$ and unsafe set $\Xu$ if there exist constants $\varepsilon_v \geq 0$ such that for all nodes $v \in \mathcal{V}$ and edges $(v,l,v') \in \mathcal{E}$ the following conditions hold:%
	\begin{subequations}\label{eq:safety-mimp}
		\begin{alignat}{2}
			\B_v(x) \leq{} & 0, \quad & \forall  &x \in \Xo, \label{eq:safety-mimp-initialSet}\\
			\B_v(x) >{} & 0, \quad & \forall &x \in \Xu, \label{eq:safety-mimp-unsafeSet}\\
			\quad \B_v(x) \leq{}& 0 \quad \Rightarrow \quad  & \forall &x \in \X, \nonumber \\
			\B_{v'}(f_{\mathrm{o}_q}^m&(\fc(x))) \leq -(l{-}m)\varepsilon_{v'} \qquad && m \in \{0, \ldots, l\}, \label{eq:safety-mimp-nodeSwitching} \raisetag{1.5\baselineskip}
		\end{alignat}%
	\end{subequations}
	where $q \in \{z,h\}$, depending on the actuator strategy used.
\end{definition}

Our first main result for safety verification is summarized below.

\begin{theorem} \label{thm:safety-mimp}
	For a WH control system under a WH constraint \anyWHRTC{r}{s} with the corresponding WH graph $\G$, suppose there exists a GBF as in Definition~\ref{def:graph-based-BF}.
	Then, the system is safe w.r.t.\ $\Xo$ and $\Xu$.
\end{theorem}

\begin{IEEEproof}
	Consider an arbitrary state sequence $\sequ{x}_{g}^{x_0}$ starting from an initial state $x_0 \in \Xo$ under the state feedback controller $g$ and loss sequence $\sequ{\mu} = \sequ{\mu}_0' \sequ{\mu}_1' \ldots$ satisfying \anyWHRTC{r}{s} with WH graph $\G$.
	For the sake of contradiction, suppose that there exists a time $T \in \N$ such that the system is unsafe, \ie $\sequ{x}_{g}^{x_0}(T) \in \Xu$.
	Moreover, suppose that this is the result of finitely many loss subsequences $\sequ{\mu}_0' \sequ{\mu}_1' \ldots \sequ{\mu}_k'$ for some $k \in \N$.
	Due to~\eqref{eq:safety-mimp-initialSet}, one has at the initial state $x_0$ and some initial node $v^0 \in \mathcal{V}$ that $\B_{v^0}(x_0) \leq 0$.
	Now consider the subsequence $\sequ{\mu}_0$ of length $l^0 + 1$ and a corresponding edge $(v^0,l^0,v^1) \in \mathcal{E}$ of $\G$ for some $l^0 \in \Sigma$.
	Then, by condition~\eqref{eq:safety-mimp-nodeSwitching}, one has that $\B_{v^1}(\sequ{x}_{g}^{x_0}(1)) = \B_{v'}(f_c(x_0)) \leq -l\varepsilon_{v'}$, $\B_{v^1}(\sequ{x}_{g}^{x_0}(2)) \leq -(l-1)\varepsilon_{v^1},\ldots,  \B_{v^1}(\sequ{x}_{g}^{x_0}(l^0+1 )) \leq 0$.
	Now, starting from node $v^1$, we consider the subsequence $\sequ{\mu}_1$ of length $l^1 + 1$ and the corresponding edge $(v^1, l^1, v^2)$, and we obtain $\B_{v^2}(\sequ{x}_{g}^{x_0}(l^0 + l^1 + 2)) \leq 0$.
	By inductively repeating this upto the subsequence $\sequ{\mu}_k$, one has that $\B_{v^{k+1}}(\sequ{x}_{g}^{x_0}(T)) \leq 0$.
	This is a contradiction to condition~\eqref{eq:safety-mimp-unsafeSet}, and therefore the WH control system remains safe.
	Note that this proof is valid for systems using both zero (as in~\eqref{eq:sys-zero}) as well as hold (as in~\eqref{eq:sys-hold1}) strategies.
\end{IEEEproof}

A couple of remarks are in order.

\begin{remark} \label{rem:hold}
	Note that despite the history dependence in $f_{\mathrm{o}_h}$ via~\eqref{eq:sys-hold1}, the tractability of condition~\eqref{eq:safety-mimp-nodeSwitching} is not affected as switching between $\B_v$ and $\B_{v'}$ occurs only at a successful control attempts.
	Therefore, the controller $g(x)$, and thus $f_c(x)$ in~\eqref{eq:safety-mimp-nodeSwitching} remains accessible for the computation of $\B_{v'}$.
\end{remark}

\begin{remark}\label{rk:eps-in-GBF}
	As a matter of fact, the least conservative version of Theorem~\ref{thm:safety-mimp} is given for $\varepsilon_v = 0$ $\forall v \in \mathcal{V}$, as the most flexibility is granted to the state evolution during losses.
	We chose to formulate \eqref{eq:safety-mimp} with general $\varepsilon_v \geq 0$ because it clarifies the behavior of the system when a loss is encountered and plays a role in our later reformulations of conditions~\eqref{eq:safety-mimp}, \cf conditions \eqref{eq:safety-1step-decrease}.
\end{remark}

\begin{remark}
	There exists a partial ordering of WH constraints, that can classify one WH constraint as \emph{harder} or \emph{weaker} than another \cite[Definition~10]{Bernat2001a}.
	A harder constraint poses more strict conditions on the number of successes.
	Any safety guarantee given in this article (verification and synthesis) that holds for a WH control system with one WH constraint also extends to any harder WH constraint.
\end{remark}

\begin{remark}
	The notion of barrier functions defined for WH control systems is closely related to $k$-inductive barrier functions presented in~\cite{Anand2021}.
	In particular, the formulation in~\cite{Anand2021} allows the barrier function to take large values at $k$ intermediate time steps as long as the zero level set is not crossed.
	Here, $k$ is a user-defined hyperparameter that allows greater computation flexibility.
	However, in our framework, we seek to utilize edge information from the WH graph to allow for flexible evaluation of barrier functions following control losses via~\eqref{eq:safety-mimp-nodeSwitching}.
	This enables dynamic, adaptive, relaxation of barrier function conditions based on future loss predictions.
\end{remark}

\begin{continuance}{ex:runningEx}[continued]
	Consider the loss sequence $\sequ{\mu} = 1 \, 0 \, 0 \, 1 \, 1 \, 0  \ldots$ corresponding to the path $v_1 2 v_3 0 v_1 1 v_2 \ldots$ that is encoded by the graph $\G$ in Figure~\ref{fig:WHgraph-example}.
	At time $t=0$ and the corresponding initial state $x_0$ and initial node $v_1$ in $\G$, equation~\eqref{eq:safety-mimp-initialSet} guarantees that $\B_{v_1}(x_0) \leq 0$.
	The first success followed by two subsequent losses triggers the transition corresponding to $(v_1,2,v_3) \in \mathcal{E}$.
	The first success ensures that the value of the barrier function at the incoming node $v_3$ is sufficiently negative and away from its zero-level set, \ie $\B_{v_3}(\sequ{x}^{x_0}_g(1)) \leq -2\varepsilon_{v_3}$.
	Now, with the two subsequent losses and in the absence of a viable actuation, $\B_{v_3}$ may move toward the unsafe set, taking up values up to $\B_{v_3}(\sequ{x}^{x_0}_g(2)) \leq -\varepsilon_{v_3}$ and $\B_{v_3}(\sequ{x}^{x_0}_g(3)) \leq 0$, respectively.
	However, since the values are still negative, the state remains within the zero-sublevel set of $\B_{v_3}$, guaranteeing safety despite losses.
	Note that the level set $\B_{v_3}(x) = 0$ can be much closer to $\Xu$ compared to $\B_{v_1}(x) = 0$, since in $v_3$ it is guaranteed that two consecutive successes will follow.
	An illustrative interplay between the GBF and the lossy state-sequence is depicted in Figure~\ref{fig:running-example-barrier}.
	\begin{figure}
		\centering
		\begin{tikzpicture}
	\usetikzlibrary{decorations.text}
	\colorlet{blue-light}{blue!60!white}
	
	% Xu
	\draw[white,fill=red,fill opacity=0.27] (-4.3,-2.9) rectangle (4.3,2.9);
	\node[red] (Xu) at (-3.5,-2.2) {$\Xu$};
	\draw[red,thick,fill=white] (0,0) ellipse (4.1 and 2.7);
	
	% Xo
	\draw[gray,thick,fill=gray,fill opacity=0.27] (0,0) ellipse (0.8 and 1);
	\node[gray] (Xo) at (0.0,0.0) {$\Xo$};
	
	% Barrier function level sets
	\draw[very thick,blue-light] (0,0) ellipse (3.0 and 2.3);
	\node[blue-light,rotate=-17] (Bv3) at (1.2,2.3) {\scriptsize $\B_{v_3} \!=\! 0$};
	
	\draw[very thick,istblue] (0,0) ellipse (1.2 and 1.3);
	\node[istblue,rotate=-17] (Bv1) at (0.5,1.355) {\scriptsize $\B_{v_{1}} \!=\! 0$};
	
	\draw[thick,blue-light,dashed] (0,0) ellipse (1.6 and 0.8);
	\node[blue-light,rotate=-30] (Bv3-2eps) at (1.75,0.4) {\scriptsize $\B_{v_3} \!=\! -2\varepsilon_{v_3}$};
	
	\draw[thick,blue-light,dashed] (0,0) ellipse (2.5 and 1.75);
	\node[blue-light,rotate=-17] (Bv3-1eps) at (1.0,1.75) {\scriptsize $\B_{v_3} \!=\! -\varepsilon_{v_3}$};
	
	% Trajectory
	\node[smallCircle] (x0) at (0,0.9) {};
	\node[smallCircle] (x1) at (-1.4,0.1) {};
	\node[smallCircle] (x2) at (-1.8,-0.8) {};
	\node[smallCircle] (x3) at (-1.6,-1.8) {};
	\node[smallCircle] (x4) at (-0.1,-1.1) {};
	\node[anchor=north,xshift=0.5em,yshift=-0.1em] (x0label) at (x0) {\small $x_0$};
	\node[anchor=east,xshift=0.2em,yshift=0em] (x1label) at (x1) {\small $\sequ{x}(1)$};
	\node[anchor=east,xshift=0.2em,yshift=0em] (x2label) at (x2) {\small $\sequ{x}(2)$};
	\node[anchor=north west,xshift=0em,yshift=0.55em] (x3label) at (x3) {\small $\sequ{x}(3)$};
	\node[anchor=south,xshift=0em,yshift=-0.2em] (x4label) at (x4) {};
	
	% arrows between trajectory points
	\draw (x0) edge[thick,dotted,-to] (x1);%node[pos=0.9,above left,xshift=3.0,yshift=0.0] {$\footnotesize 1$}
	\draw (x1) edge[thick,dotted,-to] (x2);%node[pos=0.6,below] {$\footnotesize 0$}
	\draw (x2) edge[thick,dotted,-to] (x3);%node[pos=0.4,above,xshift=-0.1,yshift=-1.0] {$\footnotesize 0$}
	\draw (x3) edge[thick,dotted,-to] (x4);%node[pos=0.6,right,xshift=0.0,yshift=0.0] {$\footnotesize 1$}
	\draw (x4) edge[thick,dotted,-to] ++(0.5,0);

\end{tikzpicture}
		\caption{An illustration of the interplay of barrier function behavior for WH constraint shown in Figure~\ref{fig:WHgraph-example}. Barrier function is steered to be sufficiently negative in the presence of success, offering flexibility via GBF to ensure safety during the following losses.
		State sequences ({\protect\tikz[baseline=-0.5ex] \protect\draw[dotted,thick] (0,0)--(0.45,0);}), initial sets (grey) and unsafe sets (red) are shown for Example~\ref{ex:runningEx}. The state dependence of $\B_{v_i}$ is omitted for ease of presentation.}
		\label{fig:running-example-barrier}
	\end{figure}
	One notices that, at $t=2$, the system state need not be within the set $\B_{v_1}(x) \leq 0$ as long as it is within $\B_{v_3}(x) \leq 0$.
	When the next success occurs at $t=3$, thus triggering a transition to node $v_1$ again, the state once again retreats from the unsafe set and ends up within $\B_{v_1}(x) \leq 0$.
\end{continuance}

While conditions~\eqref{eq:safety-mimp} provide a general formulation for safety verification, they may present challenges in the computational tractability.
One challenge is that~\eqref{eq:safety-mimp-nodeSwitching} requires the recursive computation of $f^m_{\mathrm{o}_{q}}$ which may be difficult even for simple linear systems as the allowable consecutive losses $l$ for a WH constraint and thus $m$ grow large.
In the following, we propose a new variant of conditions~\eqref{eq:safety-mimp} that eliminates the need for recursive computation at the cost of conservatism.

\begin{definition}[1-GBF: Zero Strategy] \label{def:1step-BF}
	Consider a WH control system employing the zero strategy (as in~\eqref{eq:sys-zero}), with the WH graph $\G = (\mathcal{V},\mathcal{E})$ and $g(x) \in \U \enspace \forall x \in \X$.
	A set of $n_\mathcal{V}$ functions $\B_v(x)\colon \X \rightarrow \mathbb{R}$ is called a \emph{1-step GBF for the zero strategy} (1-GBF-zero) for $\G$ w.r.t.\ an initial set $\Xo$ and unsafe set $\Xu$ if there exist constants $\varepsilon_v \geq 0$ such that for all nodes $v \in \mathcal{V}$ and edges $(v,l,v') \in \mathcal{E}$ conditions \eqref{eq:safety-mimp-initialSet}, \eqref{eq:safety-mimp-unsafeSet} and the following hold:%
	\begin{subequations}\label{eq:safety-1step}
		\begin{alignat}{2}
			\B_v(x) & \leq{} 0  \Rightarrow{} \B_{v'}(\fc(x)) \leq -l \varepsilon_{v'},  \quad &&\forall x \in \X, \label{eq:safety-1step-nodeSwitching} \\
			\B_{v'}(x&)  \leq{} -m \varepsilon_{v'} \Rightarrow{} \quad &&\forall x \in \X, \quad \nonumber \\
			&\B_{v'}(f_{\mathrm{o}_{z}}(x)) \leq -(m{-}1) \varepsilon_{v'}, \quad &&\hspace{1em} m \in \{1, \ldots, l\}. \label{eq:safety-1step-boundedIncrease} \raisetag{1.5\baselineskip}
		\end{alignat} %
	\end{subequations}
\end{definition}
Note that comparing Definitions~\ref{def:graph-based-BF} and~\ref{def:1step-BF}, condition~\eqref{eq:safety-mimp-nodeSwitching} is partitioned into the two conditions~\eqref{eq:safety-1step-nodeSwitching} and~\eqref{eq:safety-1step-boundedIncrease}.
However, observe that these conditions are not directly applicable to WH control systems operating with hold strategy, when represented as in~\eqref{eq:sys-hold1}.
Following Remark~\ref{rem:hold}, one notices that the values of $\B_{v'}$ in condition~\eqref{eq:safety-1step-boundedIncrease} depend on the system states that follow after losses, which are in turn dependent on a past state evaluated from condition~\eqref{eq:safety-1step-nodeSwitching} and are no longer accessible.
In order to alleviate this issue, we utilize the augmented state representation of the system~\eqref{eq:sys-hold2} and redefine the safety conditions for the hold strategy as follows.

\begin{definition}[1-GBF: Hold Strategy] \label{def:1step-BF-hold}
	Consider a WH control system employing the hold strategy (as in~\eqref{eq:sys-hold2}), with the WH graph $\G = (\mathcal{V},\mathcal{E})$ and $g(x) \in \U \enspace \forall x \in \X$.
	A set of $n_\mathcal{V}$ functions $\B_v(\tilde{x})\colon \X \times \U \rightarrow \mathbb{R}$ is called \emph{1-step GBF for the hold strategy} (1-GBF-hold) for $\G$ w.r.t.\ an initial set $\Xo$ and unsafe set $\Xu$, if there exist constants $\varepsilon_v \geq 0$ such that for all nodes $v \in \mathcal{V}$ and edges $(v,l,v') \in \mathcal{E}$ the following conditions hold:%
	\begin{subequations}\label{eq:safety-1step-hold}
		\begin{alignat}{2}
			& \B_v(\tilde x) \leq 0, \quad && \forall \tilde{x} \in \{ \Xo \times \mathcal{U} \mid \ua_h = g(x) \}, \label{eq:safety-1step-initialSet-hold} \\
			& \B_v(\tilde x) > 0, \quad && \forall \tilde x \in \Xu \times \U, \label{eq:safety-1step-unsafeSet-hold} \\
			& \B_v(\tilde{x}) \leq 0 \Rightarrow \notag \\
			&\hspace{1em}\B_{v'}(\tilde f_{\mathrm{c}_h}(\tilde{x})) \leq -l \varepsilon_{v'}, \quad && \forall \tilde{x} \in \X \times \mathcal{U}, \label{eq:safety-1step-nodeSwitching-hold}
				\raisetag{1.5\baselineskip} \\
			& \B_{v'}(\tilde{x}) \leq -m \varepsilon_{v'} \Rightarrow  && \forall \tilde x \in \X \times \U, \notag \\
			&\hspace{1em} \B_{v'}(\tilde f_{\mathrm{o}_h}(\tilde{x})) \leq -(m{-}1)\varepsilon_{v'},
			&&\hspace{1em} m \in \{1, \ldots, l\},
			\hspace{2em}
			\label{eq:safety-1step-boundedIncrease-hold}
			\raisetag{1.5\baselineskip}
		\end{alignat}
	\end{subequations}
	where $\tilde x = \begin{bmatrix} x & \ua_h \end{bmatrix}$ is the augmented state of the system~\eqref{eq:sys-hold2}.
\end{definition}

The following corollary results from the application of Definitions~\ref{def:1step-BF} and~\ref{def:1step-BF-hold} for WH control systems with zero and hold strategies, respectively.

\begin{corollary} \label{cor:safety-1step}
	For a WH control system under a WH constraint $\anyWHRTC{r}{s}$ with the corresponding WH graph $\G$ employing the zero or hold strategy, suppose there exists a 1-GBF-zero as in Definition~\ref{def:1step-BF} or a 1-GBF-hold as in Definition~\ref{def:1step-BF-hold}.
	Then, the system is safe w.r.t.\ $\Xo$ and $\Xu$.
\end{corollary}

\begin{IEEEproof}
	For a WH control systems employing the zero strategy, we show that the satisfaction of \eqref{eq:safety-1step} implies~\eqref{eq:safety-mimp-nodeSwitching}.
	Consider an edge $(v,l,v') \in \E$ of $\G$, and corresponding barrier functions $\B_v$ and $\B_{v'}$, for nodes $v$ and $v'$, respectively.
	From~\eqref{eq:safety-1step-nodeSwitching}, we directly arrive at~\eqref{eq:safety-mimp-nodeSwitching} for $m=0$.
	By recursively applying~\eqref{eq:safety-1step-boundedIncrease} for $m \in \{1, \ldots, l\}$, the right hand side of~\eqref{eq:safety-mimp-nodeSwitching} is obtained and by Theorem~\ref{thm:safety-mimp}, safety follows.
	
	For a WH control system using the hold strategy, conditions~\eqref{eq:safety-mimp} should be valid regardless of the input values held during the evolution of the system.
	Therefore, when utilizing the augmented state $\tilde x = \begin{bmatrix} x & \ua_h \end{bmatrix}$, conditions~\eqref{eq:safety-1step-unsafeSet-hold}--\eqref{eq:safety-1step-boundedIncrease-hold} are imposed on all possible input values in $\U$.
	Note that since the first time instant is assumed to be a success, such restriction is not imposed for~\eqref{eq:safety-1step-initialSet-hold}.
	The rest of the proof follows similarly from that of the zero strategy.
\end{IEEEproof}

Note that conditions~\eqref{eq:safety-1step} and~\eqref{eq:safety-1step-hold} are at least as conservative as conditions~\eqref{eq:safety-mimp}, since conditions~\eqref{eq:safety-1step-boundedIncrease} and~\eqref{eq:safety-1step-boundedIncrease-hold} enforce a transitive implication between the values of $\B_{v'}$ at states resulting from allowable losses, while no such restriction is placed via condition~\eqref{eq:safety-mimp-nodeSwitching}.
However, they offer the benefit of not requiring recursive computation of the dynamics.
Note further that the aforementioned conditions use logical implications, preventing one from using tools like sum-of-squares (SOS) optimization, which generally require conditions to represented in conjunctive form~\cite{Anand2021}.
To alleviate this, one can use S-procedure with a multiplier fixed to identity, and reformulate conditions~\eqref{eq:safety-1step-nodeSwitching}--\eqref{eq:safety-1step-boundedIncrease} to obtain the so-called \emph{$1$-step decrease GBF} for the zero strategy (1d-GBF-zero) with $\varepsilon_v > 0$, $\forall v \in \V$, as follows:%
\begin{subequations}\label{eq:safety-1step-decrease}
	\begin{alignat}{2}
		\B_{v'}(\fc(x)) - \B_v(x) &\leq -l\varepsilon_{v'}, \quad &&\forall x \in \X, \label{eq:safety-1step-decrease-nodeSwitching} \\
		\B_{v'}(f_{\mathrm{o}_{z}}(x)) - \B_{v'}(x) &\leq{} \varepsilon_{v'}, \quad &&\forall x \in \X \text{ if $l \geq 1$}. \label{eq:safety-1step-decrease-boundedIncrease}
	\end{alignat}
\end{subequations}
The same can be done for the hold strategy by reformulating conditions~\eqref{eq:safety-1step-nodeSwitching-hold}--\eqref{eq:safety-1step-boundedIncrease-hold} to obtain a \emph{$1$-step decrease GBF for the hold strategy} (1d-GBF-hold) as follows:
\begin{subequations}\label{eq:safety-1step-decrease-hold}
	\begin{alignat}{2}
		\B_{v'}(\fc(\tilde{x})) - \B_v(\tilde{x}) &\leq -l\varepsilon_{v'}, \quad &&\forall \tilde{x} \in \X \times \U, \label{eq:safety-1step-decrease-nodeSwitching-hold} \\
		\B_{v'}(f_{\mathrm{o}_h}(\tilde{x})) - \B_{v'}(\tilde{x}) &\leq{} \varepsilon_{v'}, \quad &&\forall \tilde{x} \in \X \times \U \text{ if $l \geq 1$}. \label{eq:safety-1step-decrease-boundedIncrease-hold}
	\end{alignat}
\end{subequations}

Here, the logical implication in conditions~\eqref{eq:safety-1step-nodeSwitching} and~\eqref{eq:safety-1step-boundedIncrease} or~\eqref{eq:safety-1step-nodeSwitching-hold} and~\eqref{eq:safety-1step-boundedIncrease-hold} are replaced by decrease conditions~\eqref{eq:safety-1step-decrease-nodeSwitching} and~\eqref{eq:safety-1step-decrease-boundedIncrease} or~\eqref{eq:safety-1step-decrease-nodeSwitching-hold} and~\eqref{eq:safety-1step-decrease-boundedIncrease-hold}, respectively.
In the former case, negativity of the consequents (i.e., values $\B_{v'}$, as regulated by $m$ and $l$) is only enforced when the antecedents are themselves sufficiently negative.
However, in the latter, the values of $\B_{v'}$ must decrease irrespective of the values taken at the previous time instant.
This unconditional decrease requirement introduces additional conservatism, but nevertheless guarantees safety, as $\B_{v'}$ is still ensured to be negative as the WH control system evolves under losses.
For more intuitive description on the difference between logical implication-based conditions and decrease conditions, we refer the reader to~\cite{Anand2021}, which presented similar conditions in the context of simple discrete-time controls systems.

Conditions~\eqref{eq:safety-1step-decrease} eliminate the need for both logical implications as well as the recursive computation of the dynamics $f_{\mathrm{o}_{q}}^m$.
However, when considering WH constraints with small number of allowable losses, it may be unnecessary to consider the more conservative formulation in~\eqref{eq:safety-1step-decrease} or~\eqref{eq:safety-1step-decrease-hold}, respectively. %
In such cases, one may still allow for a recursive computation by adapting Definition~\ref{def:graph-based-BF} using the $S$-procedure, setting $\varepsilon_v > 0$, and replace~\eqref{eq:safety-mimp-nodeSwitching} by
\begin{align} \label{eq:safety-mstep-decrease}
	\begin{split}
		\B_{v'}(f_{\mathrm{o}_{q}}^m(\fc(x))) -{}& \B_v(x) \leq -(l{-}m) \varepsilon_{v'},\\
		&\forall x \in \X, \hspace{0.5em} m \in \{0, \ldots, l\}.
	\end{split}
\end{align}
We call the resulting barrier function \emph{decrease GBF} (d-GBF).
Those conditions are valid for both the zero and the hold strategy, since the recursive computation of the dynamics in~\eqref{eq:safety-mstep-decrease} enables to keep track of the last held control input.

The safety verification problem studied above relies on the assumption that a controller is given, \ie $g$ is known.
But generally, it is also of interest to synthesize a suitable controller function $g$ that guarantees that the control system is safe even in the presence of WH constrained losses.
The following subsection tackles this problem.

\subsection{Safe controller synthesis for WH Control Systems} \label{subsec:safecon}
Here, we address Problem~\ref{prob:safetySynthesis} by establishing a GBF-based methodology for synthesizing controllers that render a WH control system safe.
In particular, we adapt the GBF conditions of the previous section for synthesis by computing $g$ alongside the GBF.
We first present the following theorem that directly utilizes the general formulation of graph-based barrier functions as in Definition~\ref{def:graph-based-BF} for controller synthesis for WH systems employing the zero strategy.

\begin{corollary} \label{thm:safety-synth-zero}
	Consider a WH control system employing the zero strategy under the WH constraint $\anyWHRTC{r}{s}$ with the corresponding WH graph $\G$.
	If one can find a suitable controller $g$ with $g(x) \in \U \enspace \forall x \in \X$ along with a corresponding GBF as in Definition~\ref{def:graph-based-BF},
	it is guaranteed that $g$ renders the system safe w.r.t.\ $\Xo$ and $\Xu$.
\end{corollary}

\begin{remark}
	Controller synthesis using 1\nobreakdash-GBF-zero via Definition~\ref{def:1step-BF} and Corollary~\ref{cor:safety-1step}, 1d-GBF-zero \eqref{eq:safety-1step-decrease}, or d-GBF-zero \eqref{eq:safety-mstep-decrease} can be performed similarly.
\end{remark}
The proof follows directly from that of Theorem~\ref{thm:safety-mimp} by applying the designed controller $g$.
While the safe controller synthesis result follows directly from that of verification, the main challenge is the computation of such a controller.
In particular, note that when $g$ is unknown, condition~\eqref{eq:safety-mimp-nodeSwitching} becomes non-convex as $\B_{v'}$ depends on $f_c$ which in turn depends on an unknown quantity $g$.
The computation of GBFs and the corresponding controllers by adapting existing techniques to specific classes of WH control systems is discussed in Section~\ref{sec:computation}.

Moreover, one observes that Corollary~\ref{thm:safety-synth-zero} cannot be directly applied for WH control systems using the hold strategy.
This is due to the fact that~\eqref{eq:sys-hold2} is memory-dependent and requires a repeated composition of $g$ in order to obtain $f^m_{\mathrm{o}_h}$, resulting in non-convexity when finding a suitable $g$.
To resolve this, one needs to consider the augmented system formulation~\eqref{eq:sys-hold2}.
We utilize the following corollary based on 1-GBF-hold according to Definition~\ref{def:1step-BF-hold} and 1d-GBF-hold~\eqref{eq:safety-1step-decrease-hold}.

\begin{corollary} \label{thm:safety-synth-hold}
	Consider a WH control system employing the hold strategy under the WH constraint $\anyWHRTC{r}{s}$ with the corresponding WH graph $\G$.
	If one can find a suitable controller $g$  with$g(x) \in \U \enspace \forall x \in \X$ along with a corresponding 1-GBF-hold as in Definition~\ref{def:1step-BF-hold} or 1d-GBF-hold according to~\eqref{eq:safety-1step-decrease-hold}, it is guaranteed that  $g$ renders the system safe w.r.t.\ $\Xo$ and $\Xu$.
\end{corollary}
The proof follows directly from that of Corollary~\ref{cor:safety-1step} by applying the designed controller $g$.

\begin{remark}
	Theorems~\ref{thm:safety-synth-zero} and~\ref{thm:safety-synth-hold} synthesize a common controller $g$ for all nodes $v \in \V$ of the WH graph $\G$.
	It is also possible to assign a different controller for each node of $\G$ to obtain a hybrid control strategy.
	Though this has been previously explored in the context of stability of switched~\cite{ElFarra2001, Daafouz2002} and WH control systems~\cite{Linsenmayer2017,Seidel2024b}, we leave such considerations in the context of safety for future work.
\end{remark}

\section{Computation of GBFs} \label{sec:computation}
In the previous section, we presented multiple formulations of GBFs for the safety verification as well as controller synthesis for WH control systems employing both zero and hold strategies.
The most general formulation is via conditions presented in~\eqref{eq:safety-mimp}.
However, it also poses a computational challenge due to the presence of logical implication-based conditions as well as recursive computation of the dynamics.
On the other hand, 1-GBF-zero~\eqref{eq:safety-1step} and~1-GBF-hold~\eqref{eq:safety-1step-hold} as well as 1d-GBF-zero~\eqref{eq:safety-1step-decrease} and 1d-GBF-hold~\eqref{eq:safety-1step-decrease-hold}, pose less restrictions on the computational complexity at the cost of higher conservatism.
In this section, we offer a brief discussion on the computation of GBFs corresponding to the aforementioned formulations.

The numerical computation of GBFs is a challenging problem and there exists no unifying framework for their computation.
For instance, to find GBFs corresponding to Definition~\ref{def:graph-based-BF}, one needs to resolve the logical implication in~\eqref{eq:safety-mimp-nodeSwitching} via the $S$-procedure through a Lagrangian multiplier, which results in a set of non-convex constraints due to the multiplication of unknown variables in $\B{v'}$ and the corresponding Lagrangian.
Nevertheless, depending on the system dynamics (e.g., linear, polynomial, etc.), parameterization of the sets (e.g. quadratic, polynomial, etc.), as well as the GBF formulations (e.g., with or without logical implication), one may leverage existing tools (e.g., linear and bilinear matrix inequalities (LMIs, BMIs)~\cite{Boyd2004}, sum-of-squares (SOS) programming~\cite{Parrilo2003}, satisfiability modulo theory (SMT) solvers~\cite{DeMoura2011}, etc.) and adapt them accordingly to search for suitable GBFs and corresponding controllers.
For general nonlinear systems, learning-based approaches may also be used~\cite{Anand2023}.
In the following, we briefly discuss the computation of GBFs for WH control systems with linear as well as polynomial dynamics for verification and synthesis via LMIs and SOS programming, respectively, and provide some insight into existing computational frameworks that can be adapted for general nonlinear systems.

\subsection{Linear systems} \label{sec:computation-linear}
A particularly useful system class are linear WH control systems, where one can obtain suitable barrier functions and controllers by leveraging matrix inequalities.
For such systems, we have $f(\sequ{x}(t),\sequ{u}^\mathrm{a}_q(t)) = A\sequ{x}(t) + B\sequ{u}^\mathrm{a}_q(t)$ and $g(\sequ{x}(t)) = K \sequ{x}(t)$, where $A \in \mathbb{R}^{n \times n}$, $B \in \mathbb{R}^{n \times n_u}$, and $K \in \mathbb{R}^{n_u \times n}$.
In this case, it is also natural to assume quadratic set representations.
For example, the state set may be defined as $\X  = \{ x \in \mathbb{R}^n \mid \begin{bmatrix} x \\ 1\end{bmatrix}^{\!\!\top} \!\!S \begin{bmatrix} x \\ 1\end{bmatrix} \geq 0 \},$ where  $S \in \mathbb{R}^{(n+1) \times (n+1)}$ is a symmetric matrix.
Similarly, $\Xo$ and $\Xu$ are appropriately defined with symmetric matrices $S_0$ and $S_u$, respectively.
Note that such representations can define hyperrectangles, ellipsoids, polytopes, zonotopes, etc.~\cite{fazlyab2022}.
Now, by considering quadratic barrier functions of the form $\B_v(x) = \begin{bmatrix} x \\ 1\end{bmatrix}^{\!\!\top} \!\!P_{v} \begin{bmatrix} x \\ 1\end{bmatrix}$ with symmetric $P_{v} \in \mathbb{R}^{(n+1) \times (n+1)}$ for all $v \in \mathcal{V}$, the required safety conditions via GBFs can be reformulated as suitable matrix inequalities.
Particularly, following techniques from~\cite{Anand2024a}, one can easily formulate all the presented graph-based safety conditions into a set of matrix inequalities via the $S$-procedure~\cite{Boyd2004}.
For example, considering GBFs of Definition~\ref{def:graph-based-BF}, conditions~\eqref{eq:safety-mimp-initialSet} and~\eqref{eq:safety-mimp-unsafeSet} may be reformulated as
\begin{equation}
	\begin{aligned}
		\alpha_v S_0 + P_v &\preccurlyeq 0 \\
		\beta_v S_\mathrm{u} - P_v &\prec 0
	\end{aligned}
\end{equation}
for all $v \in \mathcal{V}$ with multipliers $\alpha_v$, $\beta_v > 0$.
Similarly, condition~\eqref{eq:safety-mimp-nodeSwitching} results in
\begin{equation} \label{eq:BMI-safety}
	- \delta_{e,m} S + \gamma_e P_v - F^\top P_{v'} F + \begin{pmatrix}\mathbf{0} & \mathbf{0} \\ \mathbf{0} & -(l-m) \varepsilon_{v'}\end{pmatrix} \succcurlyeq 0
\end{equation}
for each edge $e = (v,l,v') \in \mathcal{E}$ and all $m \in \{0, \ldots, l\}$, where $\delta_{e,m}$, $\gamma_e > 0$ are multipliers, $F = \begin{pmatrix}A^m (A+BK) & \mathbf{0} \\ \mathbf{0} & \mathbf{0} \end{pmatrix}$, and $\mathbf{0}$ denotes a zero matrix of appropriate dimension. Conditions~\eqref{eq:safety-1step}--\eqref{eq:safety-mstep-decrease} may be reformulated analogously.
However, note that due to the presence of logical implication, condition~\eqref{eq:safety-mimp-nodeSwitching} from Definition~\ref{def:graph-based-BF}, conditions~\eqref{eq:safety-1step-nodeSwitching}--\eqref{eq:safety-1step-boundedIncrease} and~\eqref{eq:safety-1step-nodeSwitching-hold}--\eqref{eq:safety-1step-boundedIncrease-hold} from Definitions~\ref{def:1step-BF} and~\ref{def:1step-BF-hold}, respectively, result in BMIs, while conditions~\eqref{eq:safety-1step-decrease-nodeSwitching}--\eqref{eq:safety-1step-decrease-boundedIncrease}, and~\eqref{eq:safety-1step-decrease-nodeSwitching-hold}--\eqref{eq:safety-1step-decrease-boundedIncrease-hold}, respectively, i.e., the ones without logical implication, result in simpler LMIs.
For example, in the reformulation of condition~\eqref{eq:safety-mimp-nodeSwitching} presented in~\eqref{eq:BMI-safety}, bilinearity stems from the term $\gamma_e P_v$.
Nevertheless, one may utilize a bisection approach to find a feasible solution for the BMIs by solving LMIs in an iterative manner until convergence to a valid solution is achieved \cite{VanAntwerp2000}.

Now we describe a solution to the controller synthesis problem.
For simplicity, we first provide an approach to synthesize controllers for GBFs corresponding to conditions~\eqref{eq:safety-mimp}.
Note that in the respective condition~\eqref{eq:BMI-safety}, the controller $K$ to be designed appears quadratic in the term $F^\top P_{v'} F$.
However, it is still possible to solve~\eqref{eq:BMI-safety} as controller design problem in an iterative fashion under some mild assumptions on the structure of the barrier function.
Denote $P_v = \begin{pmatrix} p_{v,1} & p_{v,2} \\ p_{v,2}^\top & p_{v,3} \end{pmatrix}$ and suppose that $p_{v,1} \succ 0$ for all $v \in \mathcal{V}$.
One starts with an fixed initial controller $K$ and initial values of $\gamma_v > 0$ and solve~\eqref{eq:BMI-safety} as an LMI to find suitable $P_v$.
In the next iteration, $P_v$ for all $v \in \mathcal{V}$ are then fixed and~\eqref{eq:BMI-safety} is solved for $K$ and $\gamma_v$.
Due to the structure of $P_v$, $v \in \mathcal{V}$, the search for $K$ via~\eqref{eq:BMI-safety} can be recast equivalently as an LMI as follows:
Condition~\eqref{eq:BMI-safety} is equivalent to
\begin{equation}
	\begin{aligned}
		- &\delta_{e,m} S + \gamma_e P_v
		- V^\top p_{v'\!,3} V
		- V^\top p_{v'\!,2}^\top W \\
		- &W^\top p_{v'\!,2} V
		- W^\top p_{v'\!,1} W
		+ \begin{pmatrix}\mathbf{0} & \mathbf{0} \\ \mathbf{0} & -(l-m) \varepsilon_{v'}\end{pmatrix} \succcurlyeq 0,
	\end{aligned}
\end{equation}
where $W \coloneq \begin{pmatrix} A^m(A+BK) & \mathbf{0} \end{pmatrix}$ and $V \coloneq \begin{pmatrix} \mathbf{0} & 1 \end{pmatrix}$.
Then, using the Schur complement results in
\begin{equation} \label{eq:LMI-from-QMI}
	\begin{pmatrix}
		(p_{v'\!,1})^{-1} \!\! & W \\
		W^\top 			& - \delta_{e,m} S + \gamma_e P_v + (\ast) + \begin{pmatrix}\mathbf{0} & \mathbf{0} \\ \mathbf{0} & \!\!-(l-m)\end{pmatrix}
	\end{pmatrix} \succcurlyeq 0,
\end{equation}
where $(\ast) = - V^\top p_{v'\!,3} V - V^\top p_{v'\!,2}^\top W - W^\top p_{v'\!,2} V $.
Since the decision variables only appear linearly, \eqref{eq:LMI-from-QMI} constitutes an LMI.

\noindent{\textbf{Computational Complexity.}} Solving BMIs is in general an NP-hard problem~\cite{bmi_hard}, and there exists no convergence guarantees.
However, since we look for a feasible solution of BMIs through iterative programming methods, we present the per-iteration complexity of the approach using LMIs.
In general, the complexity depends on the number of nodes and edges of $\mathcal{G}$, denoted $n_{\mathcal{V}} = \abs{\mathcal{V}}$ and $n_{\mathcal{E}} = \abs{\mathcal{E}}$, respectively, which has an exponential worst case growth with the window size $s$ \cite{Vreman2022c}.
For a fixed graph size, the complexity of solving LMIs is worst case polynomial both in the number of constraints as well as number of variables~\cite{vandenberghe_semidefinite_1996}.
Note that, for standard GBF conditions~\eqref{eq:safety-mimp}, one must compute $n_{\mathcal{V}}$ matrices of dimension $(n+1) \times (n+1)$ corresponding to the GBF, consisting of $\frac{n_{\mathcal{V}}}{2}n(n+1)$ variables in total.
In addition, one has at most $3n_{\mathcal{V}} + 2kn_{\mathcal{E}}$ slack variables, where $k = s-r+1$.
Similarly, for 1-GBF-zero via conditions~\eqref{eq:safety-1step}, one has $3n_{\mathcal{V}} + 2kn_{\mathcal{E}}$ variables as well and 1d-GBF-zero~\eqref{eq:safety-1step-decrease} requires $3n_{\mathcal{V}} + 2n_{\mathcal{E}}$ slack variables.
The GBF matrix variables remain the same as for the standard case.
However, due to the augmented state construction, 1-GBF-hold via conditions~\eqref{eq:safety-1step-hold} and 1d-GBF-hold via~\eqref{eq:safety-1step-decrease-hold} require the GBF to be of size $(n+m+1) \times (n+m+1)$, thus increasing the complexity in comparison to the zero case.
Finally, note that the complexity of computing matrix powers (i.e., for computation of recursive dynamics) is logarithmic in $k$.
Therefore, for high values of $k$, standard GBF conditions~\eqref{eq:safety-mimp} has an additional logarithmic complexity w.r.t.\ $k$ that grows faster than 1-GBF and 1d-GBF conditions.

\subsection{Polynomial systems} \label{sec:computation-polynomial}
When the underlying dynamics of the WH control system are polynomial, and the initial set $\Xo$ and the unsafe set $\Xu$ are semi-algebraic sets, one may reformulate the safety conditions as an SOS optimization problem \cite{Parrilo2003} and search for a suitable graph-based polynomial barrier function of a predefined degree.
However, note that the SOS problem requires the set of constraints to be in conjunctive form, i.e., the constraints are written as a conjunction of a collection of constraints.
As a result, SOS optimization is only suitable for the computation of polynomial barrier functions satisfying conditions~\eqref{eq:safety-1step-decrease} and~\eqref{eq:safety-1step-decrease-hold}.

As an example, we reformulate condition~\eqref{eq:safety-mimp-initialSet} as an SOS condition.
Suppose that $f$ and $g$ are given by polynomials over the state $x$, $\X$ is a continuous state set, and the initial set as well as unsafe set can be described by $\Xo = \{x \in \X \mid g_0(x) \geq 0 \}$, and $\Xu = \{x \in X \mid g_\mathrm{u}(x) \geq 0 \}$, where $g_0, g_\mathrm{u}$ are polynomials.
Then, consider a polynomial 1d-GBF given by $\B_{v} = \sum_{i=1}^m c_i p_i(x)$, where $p_i$ are monomials over $x$.
Condition~\eqref{eq:safety-mimp-initialSet} may be reformulated as
	\begin{equation}
		\B_v(x) - g_0(x)\lambda_0(x) \geq 0, 
	\end{equation}
where $\lambda_0(x)$ is a polynomial Lagrangian multiplier.
Finding a suitable $\B_v$ and $\lambda_0$ that render the left hand side of the inequality as an SOS polynomial ensures the satisfaction of the inequality. Conditions~\eqref{eq:safety-mimp-unsafeSet} and \eqref{eq:safety-1step-decrease-nodeSwitching}--\eqref{eq:safety-1step-decrease-boundedIncrease} may be reformulated similarly.
For details on reformulating controller synthesis using SOS, we refer the reader to~\cite{Japtap_2021}.

\begin{remark} \label{rem:computational-complexity-polynomial}
	Similar to the linear case, the complexity of solving SOS constraints is polynomial with respect to the number of decision variables and constraints \cite{Parrilo2003}.
	Moreover, unlike LMIs, SOS problems cannot handle logical implications, which necessitates the use of d-GBFs and 1d-GBFs.
	The recursive computation used in d-GBFs additionally grows linearly w.r.t.\ the permissible number of consecutive losses $s-r$.
\end{remark}

\begin{remark}
	For the computation of GBFs that do not conform to SOS reformulations, i.e., for conditions with logical implications, e.g.,~\eqref{eq:safety-mimp}, or for non-polynomial system classes, alternate methods exist.
	For example in~\cite{Anand2021}, a satisfiability modulo theory (SMT)-based approach is proposed, which can be adapted for the computation of GBFs.
	Further, neural network parameterizations have become widely popular for data-driven computation of barrier functions~\cite{edwards_general_2025} due to their representative power beyond polynomial functions, which can be also utilized for the computation of GBFs for general nonlinear WH control systems.
	We reserve the presentation of such approaches for future work.
\end{remark}

\section{Numerical Case Studies} \label{sec:numerics}
In this section, we demonstrate the effectiveness of our results through a variety of numerical case studies spanning both safety verification and controller synthesis for WH control systems with linear as well as polynomial dynamics.

\subsection{Safety verification} \label{sec:numerics-linearVerification}
In this section, we deal with numerical examples concerning safety verification.
We present three case studies.
First, we restrict ourselves to WH control systems with linear dynamics, and demonstrate safety verification via GBFs for both zero and hold actuator strategies.
Then, we also consider an application-oriented polynomial system.

\begin{casestudy}[Hold] \label{ex:linear-verification-Hold}
	In this case study, we consider an academic linear WH control system with the following dynamics
	\begin{equation} \label{eq:academic-example}
		\sequ{x}(t+1) = \begin{bmatrix}
				0 & 1 \\
				1 & 1
			\end{bmatrix} \sequ{x}(t) + \begin{bmatrix}
				1 \\ 1
			\end{bmatrix} \sequ{u}^\mathrm{a}(t).
	\end{equation}
	which has been adapted from \cite{Blind2015}. Moreover, consider the controller $K = \begin{bmatrix}-0.5 & -0.7\end{bmatrix}$, and the WH constraint \anyWHRTC{2}{4} with WH graph depicted in Figure~\ref{fig:WHgraph-example}.
	Let the initial and the unsafe sets be
	\begin{equation}
		\begin{aligned} \label{eq:running-example-sets}
			\Xo &= \{ x \in \X \subseteq \mathbb{R}^2 \mid x_1^2 + x_2^2 \leq 0.4^2 \},\\
			\Xu &= \{ x \in \X \mid \begin{bmatrix} x \\ 1\end{bmatrix}^{\!\!\top}  \begin{bmatrix}
				-0.2 & 0 & 0.3 \\ 0 & 0 & 0.5 \\ 0.3 & 0.5 & -1
			\end{bmatrix}\begin{bmatrix} x \\ 1\end{bmatrix} \geq 0\}.
		\end{aligned}
	\end{equation}
	We use this to provide a detailed account on the computation of GBFs using different formulations, including standard ones (conditions~\eqref{eq:safety-mimp}), 1d-GBF-hold (conditions~\eqref{eq:safety-1step-hold}), and d-GBF  (condition~\eqref{eq:safety-mstep-decrease}).
	For the hold strategy, we compute the relevant GBFs (corresponding to the safety conditions mentioned above) numerically by reformulating the conditions as suitable matrix inequalities as shown in Section~\ref{sec:computation-linear} and employing the YALMIP~\cite{Lofberg2004} toolbox equipped with MOSEK~\cite{mosek} solver on MATLAB 2024a on a machine with 16\,GB RAM running Windows 11.

	We observe that safety verification is possible with the help of GBF and d-GBF-hold, respectively. That is, we are able to successfully compute $\B_{v_1}, \B_{v_2},$ and $\B_{v_3}$ for the WH graph in Figure~\ref{fig:WHgraph-example} satisfying conditions~\eqref{eq:safety-mimp} and~\eqref{eq:safety-mstep-decrease}, respectively.
	However, we fail to find a 1d-GBF-hold satisfying conditions~\eqref{eq:safety-1step-decrease-hold}.
	This is attributed to the fact that 1d-GBFs are generally more conservative than the aforementioned counterparts.
	The resulting barrier functions are depicted in Figure~\ref{fig:running-example-Hold}, along with an illustrative state sequence.
	\begin{figure}[t]
		\centering
		\input{fig/Any24Hold.tex}
		\caption{Zero-level sets of the GBF $\B$ (solid)
			and d-GBF-hold $\B'$ (dashed)
			for the WH control system of Case Study~\ref{ex:linear-verification-Hold} with the WH graph of Figure~\ref{fig:WHgraph-example}, and an example state sequence under the loss sequence $\sequ{\mu} = 1 \, 0 \, 0 \, 1 \, 1$.
		}
		\label{fig:running-example-Hold}
	\end{figure}
	Observe that after two consecutive losses, the state is not within the set $\B_{v_1} \leq 0$ nor in the set $\B_{v_2} \leq 0$, but within $\B_{v_3} \leq 0$. 
	This aligns with the original intention of the design (c.f. Figure~\ref{fig:running-example-barrier}).
\end{casestudy}

Next, we consider a different WH constraint and initial set under the zero strategy, which reveals the difference in conservatism across different formulations of GBFs.

\begin{casestudy}[Zero] \label{ex:linear-verification-Zero}
	In this case study, we again consider the academic system from~\eqref{eq:academic-example}, but operating under the WH constraint $\anyWHRTC{3}{7}$ and the zero strategy.
	The respective WH graph has 15 nodes and its illustration is thus ommitted.
	However, for demonstration, a subgraph consisting of $3$ nodes is depicted in Figure~\ref{fig:WHgraph-3-7}.
	The controller is predefined for the system~\cite{Blind2015} as $\uc(x) = Kx$, where $K = \begin{bmatrix}-0.35 & -0.85\end{bmatrix}$.
	Moreover, consider the initial set given by $\Xo = \{ x \in \mathbb{R}^2 \mid x_1^2/0.21 + x_2^2/0.5 \leq 1^2 \}$ and unsafe set as in~\eqref{eq:running-example-sets}.
	Under the zero actuator strategy, we again investigate safety for the three different GBF variants, as in Case Study~\ref{ex:linear-verification-Hold}.
	Now, similar to the previous example, we consider safety verification using standard GBFs as in~\eqref{eq:safety-mimp}, 1d-GBF-zero as in~\eqref{eq:safety-1step-decrease} and d-GBF as in~\eqref{eq:safety-mstep-decrease} by reformulating the corresponding safety conditions as matrix inequalities.
	We obtain the GBF $\B$ and d-GBF $\B'$, respectively, while the 1d-GBF-zero fails to provide feasible results.
	The initial set, an example state sequence, and the resulting GBFs for nodes $v_i \in \{v_1, v_2, v_3\}$ are depicted in Figure~\ref{fig:linear-verification-example-Zero}.
	$\B_{v_i}$ and $\B'_{v_i}$ for $v_i \in \{v_4, \dots, v_{15}\}$ are omitted for ease of illustration.
	\begin{figure}[t]
		\centering
		\input{fig/Any37Zero.tex}
		\caption{
			Zero-level sets of the GBF $\B_{v_i}$ (solid) and the d-GBF-zero $\B'_{v_i}$ (dashed) for the WH control system of Case Study~\ref{ex:linear-verification-Zero}, and an example state sequence under the loss sequence $\sequ{\mu} = 1 \, 0 \, 0 \, 0 \, 0 \, 1 \, 1$.
			Barriers for nodes $v_4$ through $v_{15}$ are omitted for ease of illustration.
		}
		\label{fig:linear-verification-example-Zero}
	\end{figure}
	Note that the sets $\B_{v_{i}} \leq 0$ for each $v_i$ contain the sets $\B'_{v_{i}} \leq 0$.
	The GBF is thus less restrictive than the d-GBF, because it provides a larger invariant set, but comes at the cost of solving a BMI with no convergence guarantee instead of an convex LMI problem.
	This increased flexibility is especially relevant when one needs to utilize pre-computed invariant sets (i.e., zero sub-level sets of the GBFs) for controller design. 
\end{casestudy}

Now, we present a case study for safety verification for a car platooning system with polynomial dynamics.

\begin{casestudy}[Polynomial System] \label{ex:nonlinear-verification}
	Consider a two-car platoon system adapted from \cite{Luppi2024} following the dynamics
	\begin{equation*}
		\begin{aligned}
			\sequ{x}(t+1) = 
			\begin{bmatrix}
				\sequ{u}^\mathrm{a}(t) \!+\! c_u \!-\! \gamma_1 \!+\! \sequ{x}_1(t) \!-\! \tau ( \beta_1 \sequ{x}_1(t) \!+\! \alpha_1 \sequ{x}_1(t)^2) \\
				\phantom{\sequ{u}^\mathrm{a}(t) \!+\!~} c_u \!-\! \gamma_2 \!+\! \sequ{x}_2(t) \!-\! \tau ( \beta_2 \sequ{x}_2(t) \!+\! \alpha_2 \sequ{x}_2(t)^2)
			\end{bmatrix}\!\!,
		\end{aligned}
	\end{equation*}
	where $\sequ{x}_1(t)$ and $\sequ{x}_2(t)$ are the velocities of vehicles 1 and 2 and with the parameters $\tau = \qty{1}{\second}$, $2 \alpha_1 = \alpha_2 = \qty{0.04}{\newton\square\second\per\kilogram\per\square\meter}$, $2 \beta_1 = \beta_2 = \qty{0.2}{\newton\second\per\kilogram\per\meter}$ and $\gamma_1 = \gamma_2 = \qty{0.005}{\newton\per\kilogram}$.
	Suppose the the second vehicle, i.e, the one corresponding to $x_2$, is the leader and the first vehicle corresponding to $x_1$ is tasked with following the second car.
	As a result, both vehicles drive with constant throttle $c_u = \qty{1.5}{\meter\per\square\second}$, however car 1 additionally adapts its acceleration through the control input $\sequ{u}^\mathrm{a}(t)$ to ensure a safe distance to the leading car 2.
	The initial set is given by $\Xo = \{\, \begin{bmatrix} x_1 \! & \! x_2 \end{bmatrix}^\top \!\! \in \X \mid x_1 \in [2,5],\, 3 \leq x_2 - x_1 \leq 4 \,\}$, the unsafe set is $\Xu = \{\, \begin{bmatrix} x_1 \! & \! x_2 \end{bmatrix}^\top \!\! \in \X \mid x_2 - x_1 \geq 0.2 \,\}$, and the state set is $\X = [0,10]^2$.
	For the WH constraint \anyWHRTC{3}{5} (corresponding WH graph see Figure~\ref{fig:WHgraph-3-5}), the platoon is controlled by the control law $\sequ{u}^\mathrm{a}(t) = -0.5 \sequ{x}_2(t)$ using the zero actuator strategy.
	\begin{figure}
		\begin{minipage}[t]{0.48\columnwidth}
			\centering
			\begin{tikzpicture}[scale=1.0]
\node[vertexFill] (v1) at (0,0) {$v_1$};
\node[vertexFill] (v5) at (1.1,1.1) {$v_2$}; %
\node[vertexFill] (v15) at (2.0,0) {$v_{3}$}; %

\draw[edge] (v1) to[loop above] node[above, pos=0.5] {\footnotesize$0$} (v1);
\draw[edge] (v1) to[          ] node[above, pos=0.5] {\footnotesize$4$} (v5);

\draw[edge] (v5) to[          ]  node[right, pos=0.5] {\footnotesize$0$} (v15);

\draw[edge] (v15) to[          ] node[above, pos=0.5] {\footnotesize$0$} (v1);

\draw[edge,dashed] (v1) to[] ($(v1)+(-0.4,-0.85)$);
\draw[edge,dashed] (v1) to[] ($(v1)+(-1.0,-0.4)$);
\draw[edge,dashed] (v1) to[] ($(v1)+(1.0,-0.4)$);
\draw[edge,dashed] ($(v1)+(0.0,-0.85)$) to[] (v1);
\draw[edge,dashed] ($(v1)+(-1.0,0.0)$) to[] (v1);
\draw[edge,dashed] ($(v1)+(0.7,-0.7)$) to[] (v1);

\end{tikzpicture}
			\caption{Subgraph of WH graph for the WH constraint \anyWHRTC{3}{7} with nodes corresponding to the loss sequence $\sequ{\mu} = 1 \, 0 \, 0 \, 0 \, 0 \, 1 \, 1$.}
			\label{fig:WHgraph-3-7}
		\end{minipage}%
		\hfill%
		\begin{minipage}[t]{0.48\columnwidth}
			\centering
			\begin{tikzpicture}

\begin{scope}[shift={(0,0)},scale=1.0]
	\node[vertexFill] (v1) at (0,0) {$v_1$};
	\node[vertexFill] (v2) at (-1.3,-0.8) {$v_2$};
	\node[vertexFill] (v3) at (0.8,-1.2) {$v_3$};
	\node[vertexFill] (v4) at (-1.3,0.4) {$v_4$};
	\node[vertexFill] (v5) at (-2.6,-0.2) {$v_5$};
	\node[vertexFill] (v6) at (-0.4,-1.2) {$v_6$};

	\draw[edge] (v1) to[loop above] node[right, pos=0.75] {\footnotesize$0$} (v1);
	\draw[edge] (v1) to[] node[below, pos=0.5] {\footnotesize$1$} (v2);
	\draw[edge] (v1) to[] node[right, pos=0.5] {\footnotesize$2$} (v3);
	
	\draw[edge] (v2) to[bend left] node[left, pos=0.5] {\footnotesize$0$} (v4);
	\draw[edge] (v2) to[] node[below, pos=0.5] {\footnotesize$1$} (v5);
	
	\draw[edge] (v3) to[] node[below, pos=0.5] {\footnotesize$0$} (v6);
	
	\draw[edge] (v4) to[] node[above, pos=0.5] {\footnotesize$0$} (v1);
	\draw[edge] (v4) to[bend left] node[right, pos=0.5] {\footnotesize$1$} (v2);
	
	\draw[edge] (v5) to[] node[above, pos=0.5] {\footnotesize$0$} (v4);
	
	\draw[edge] (v6) to[] node[right, pos=0.5] {\footnotesize$0$} (v1);
	
\end{scope}

\end{tikzpicture}
			\caption{WH graph for the WH constraint \anyWHRTC{3}{5}.}
			\label{fig:WHgraph-3-5}
		\end{minipage}
	\end{figure}
	To verify safety of this WH control system, we use SOS optimization as discussed in Section~\ref{sec:computation-polynomial} to compute 1d-GBF-zero as in conditions~\eqref{eq:safety-1step-decrease} with polynomial degree $n_\mathrm{p} = 3$.
	Using YALMIP~\cite{Lofberg2004} and the solver SeDuMi~\cite{Sturm1999}, we obtain the 1d-GBF-zero with zero-level sets depicted in Figure~\ref{fig:nonlinear}, verifying safety of the WH control system.
	Observe again that the zero-level set boundaries are not equally close to the unsafe set, essentially adjusting to the possibility of future losses according to the underlying WH constraint.
\end{casestudy}
\begin{figure*}
	\begin{minipage}[t]{0.32\textwidth}
		\centering
		\input{fig/nonlinear-verification.tex}
		\caption{
			Some zero-level sets of the 1d-GBF-zero $\B_{v_i}$ for the WH control system of Case Study~\ref{ex:nonlinear-verification}.
		}
		\label{fig:nonlinear}
	\end{minipage}%
	\hfill%
	\begin{minipage}[t]{0.32\textwidth}
		\centering
		\colorlet{blue-light1}{blue!60!white}
\colorlet{blue-light2}{blue!30!white}

\begin{tikzpicture}

\pgfplotsset{compat=1.11}

\begin{axis}[%
width=4.75cm,
height=4.75cm,
at={(0,0)},
scale only axis,
xmin=-1.0,
xmax=1.3,
ymin=-1.0,
ymax=1.3,
axis background/.style={fill=white},
xmajorgrids,
ymajorgrids,
legend style={legend cell align=left, align=left, at={(axis cs:1.6,0.1)}},
legend image post style={
	sharp plot,
	draw=\pgfkeysvalueof{/pgfplots/contour/draw color},
},
]

\draw[gray,thick,fill=gray,fill opacity=0.27] (axis cs:0,0) ellipse[x radius=0.315, y radius=0.75];
\node[gray] (Xo) at (axis cs:0.0,-0.2) {$\Xo$};

\draw[red,thick,fill=red,fill opacity=0.27]
	(-1,1.8) --
	(-0.98,1.78008) --
	(-0.979919028340081,1.78) --
	(-0.96,1.76032) --
	(-0.959673469387755,1.76) --
	(-0.94,1.74072) --
	(-0.939259259259259,1.74) --
	(-0.92,1.72128) --
	(-0.918672199170124,1.72) --
	(-0.9,1.702) --
	(-0.897907949790795,1.7) --
	(-0.88,1.68288) --
	(-0.876962025316456,1.68) --
	(-0.86,1.66392) --
	(-0.855829787234042,1.66) --
	(-0.84,1.64512) --
	(-0.83450643776824,1.64) --
	(-0.82,1.62648) --
	(-0.812987012987013,1.62) --
	(-0.8,1.608) --
	(-0.791266375545851,1.6) --
	(-0.78,1.58968) --
	(-0.769339207048458,1.58) --
	(-0.76,1.57152) --
	(-0.7472,1.56) --
	(-0.74,1.55352) --
	(-0.724843049327354,1.54) --
	(-0.72,1.53568) --
	(-0.702262443438914,1.52) --
	(-0.7,1.518) --
	(-0.68,1.50048) --
	(-0.679447004608295,1.5) --
	(-0.66,1.48312) --
	(-0.656372093023256,1.48) --
	(-0.64,1.46592) --
	(-0.633051643192488,1.46) --
	(-0.62,1.44888) --
	(-0.609478672985782,1.44) --
	(-0.6,1.432) --
	(-0.585645933014354,1.42) --
	(-0.58,1.41528) --
	(-0.561545893719807,1.4) --
	(-0.56,1.39872) --
	(-0.54,1.38232) --
	(-0.537142857142857,1.38) --
	(-0.52,1.36608) --
	(-0.512437810945273,1.36) --
	(-0.5,1.35) --
	(-0.487437185929648,1.34) --
	(-0.48,1.33408) --
	(-0.462131979695432,1.32) --
	(-0.46,1.31832) --
	(-0.44,1.30272) --
	(-0.436476683937824,1.3) --
	(-0.42,1.28728) --
	(-0.410471204188482,1.28) --
	(-0.4,1.272) --
	(-0.384126984126984,1.26) --
	(-0.38,1.25688) --
	(-0.36,1.24192) --
	(-0.357405405405405,1.24) --
	(-0.34,1.22712) --
	(-0.330273224043716,1.22) --
	(-0.32,1.21248) --
	(-0.302762430939226,1.2) --
	(-0.3,1.198) --
	(-0.28,1.18368) --
	(-0.274802259887006,1.18) --
	(-0.26,1.16952) --
	(-0.2464,1.16) --
	(-0.24,1.15552) --
	(-0.22,1.14168) --
	(-0.217543859649123,1.14) --
	(-0.2,1.128) --
	(-0.188165680473373,1.12) --
	(-0.18,1.11448) --
	(-0.16,1.10112) --
	(-0.15830303030303,1.1) --
	(-0.14,1.08792) --
	(-0.127852760736196,1.08) --
	(-0.12,1.07488) --
	(-0.1,1.062) --
	(-0.0968553459119498,1.06) --
	(-0.0800000000000001,1.04928) --
	(-0.0652229299363058,1.04) --
	(-0.0600000000000001,1.03672) --
	(-0.04,1.02432) --
	(-0.0329411764705883,1.02) --
	(-0.02,1.01208) --
	(0,1) --
	(0,1) --
	(0.02,0.98808) --
	(0.0337414965986395,0.98) --
	(0.04,0.97632) --
	(0.0600000000000001,0.96472) --
	(0.0682517482517483,0.96) --
	(0.0800000000000001,0.95328) --
	(0.1,0.942) --
	(0.103597122302158,0.94) --
	(0.12,0.93088) --
	(0.13985401459854,0.92) --
	(0.14,0.91992) --
	(0.16,0.90912) --
	(0.177142857142857,0.9) --
	(0.18,0.89848) --
	(0.2,0.888) --
	(0.215503875968992,0.88) --
	(0.22,0.87768) --
	(0.24,0.86752) --
	(0.25504,0.86) --
	(0.26,0.85752) --
	(0.28,0.84768) --
	(0.295867768595042,0.84) --
	(0.3,0.838) --
	(0.32,0.82848) --
	(0.338119658119658,0.82) --
	(0.34,0.81912) --
	(0.36,0.80992) --
	(0.38,0.80088) --
	(0.381981981981982,0.8) --
	(0.4,0.792) --
	(0.42,0.78328) --
	(0.42766355140187,0.78) --
	(0.44,0.77472) --
	(0.46,0.76632) --
	(0.475339805825243,0.76) --
	(0.48,0.75808) --
	(0.5,0.75) --
	(0.52,0.74208) --
	(0.525360824742269,0.74) --
	(0.54,0.73432) --
	(0.56,0.72672) --
	(0.578064516129033,0.72) --
	(0.58,0.71928) --
	(0.6,0.712) --
	(0.62,0.70488) --
	(0.634022988505748,0.7) --
	(0.64,0.69792) --
	(0.66,0.69112) --
	(0.68,0.68448) --
	(0.693827160493827,0.68) --
	(0.7,0.678) --
	(0.72,0.67168) --
	(0.74,0.66552) --
	(0.7584,0.66) --
	(0.76,0.65952) --
	(0.78,0.65368) --
	(0.8,0.648) --
	(0.82,0.64248) --
	(0.829253731343283,0.64) --
	(0.84,0.63712) --
	(0.86,0.63192) --
	(0.88,0.62688) --
	(0.9,0.622) --
	(0.908474576271186,0.62) --
	(0.92,0.61728) --
	(0.94,0.61272) --
	(0.96,0.60832) --
	(0.98,0.60408) --
	(1,0.6) --
	(1,0.6) --
	(1.02,0.59608) --
	(1.04,0.59232) --
	(1.06,0.58872) --
	(1.08,0.58528) --
	(1.1,0.582) --
	(1.11282051282051,0.58) --
	(1.12,0.57888) --
	(1.14,0.57592) --
	(1.16,0.57312) --
	(1.18,0.57048) --
	(1.2,0.568) --
	(1.22,0.56568) --
	(1.24,0.56352) --
	(1.26,0.56152) --
	(1.27652173913043,0.56) --
	(1.28,0.55968) --
	(1.3,0.558) --
	(1.32,0.55648) --
	(1.34,0.55512) --
	(1.36,0.55392) --
	(1.38,0.55288) --
	(1.4,0.552) --
	(1.42,0.55128) --
	(1.44,0.55072) --
	(1.46,0.55032) --
	(1.48,0.55008) --
	(1.5,0.55) --
	(2.0,2.0) --
	cycle;
\node[red] (Xu) at (0.8,0.9) {$\Xu$};

\node[smallCircle,minimum size=2pt] (x0) at (0.180676577450579, 0.614364033216744) {};
\node[smallCircle,minimum size=2pt] (x1) at (0.028917802874809, 0.209594380325388) {};
\node[smallCircle,minimum size=2pt] (x2) at (0.209594380325388, 0.238512183200197) {};
\node[smallCircle,minimum size=2pt] (x3) at (0.238512183200197, 0.448106563525586) {};
\node[smallCircle,minimum size=2pt] (x4) at (0.448106563525586, 0.686618746725783) {};
\node[smallCircle,minimum size=2pt] (x5) at (0.686618746725783, 1.134725310251368) {};

\draw (x0) edge[thick,densely dotted,-to] (x1);
\draw (x1) edge[thick,densely dashed,-to,red] (x2);
\draw (x2) edge[thick,densely dashed,-to,red] (x3);
\draw (x3) edge[thick,densely dashed,-to,red] (x4);
\draw (x4) edge[thick,densely dashed,-to,red] (x5);

\end{axis}

\end{tikzpicture}%
		\caption{
			Unsafe behavior of the WH control system of Case Study~\ref{ex:linear-synthesis-zero} shown by an example state sequence.
		}
		\label{fig:linear-synth-unsafe}
	\end{minipage}%
	\hfill%
	\begin{minipage}[t]{0.32\textwidth}
		\centering
		\input{fig/Any37ZeroSynth.tex}
		\caption{
			Safe controller synthesis of Case Study~\ref{ex:linear-synthesis-zero}: zero-level sets of the GBF $\B_{v_i}$, and an example state sequence of the system under the synthesized controller.
			Barriers $\B_{v_4}$ through $\B_{v_{15}}$ are omitted. %
		}
		\label{fig:linear-synth-example}
	\end{minipage}%
\end{figure*}

\subsection{Safe controller synthesis} \label{sec:numerics-linearSynthesis}
In Case Study~\ref{ex:linear-verification-Zero}, we verified that a WH control system is safe w.r.t.\ to the initial set $\Xo$ depicted in Figure~\ref{fig:linear-verification-example-Zero}.
Now, we aim to find a controller that renders the same system safe w.r.t.\ a larger initial set.

\begin{casestudy}[Controller Synthesis]\label{ex:linear-synthesis-zero}
	Consider the WH control system from Case Study~\ref{ex:linear-verification-Zero}, operating under the WH constraint \anyWHRTC{3}{7} with the WH graph depicted in Figure~\ref{fig:WHgraph-3-7} and zero actuator strategy.
	Now, suppose that the initial set is enlarged to $\Xo = \{ x \in \mathbb{R}^2 \mid x_1^2/0.21 + x_2^2/0.5 \leq 1.5^2 \}$ and $\Xu$ as in~\eqref{eq:running-example-sets}.
	When using the controller $K$ from Case Study~\ref{ex:linear-verification-Zero}, we observe that the system is no longer safe, as there exist state sequences that start in $\Xo$ and lead to $\Xu$ as illustrated in Figure~\ref{fig:linear-synth-unsafe}.
	As a matter of fact, when trying to verify safety using a GBF (Definition~\ref{def:graph-based-BF}), the solver of the underlying optimization problem reports infeasibility.
	We now use the approach described in Section~\ref{sec:computation-linear} to find a a linear controller $\uc(x) = K'x$ that ensures safety.
	In particular, by considering a GBF, reformulating the conditions as BMIs, and solving for a suitable $K'$, we obtain $K' = \begin{bmatrix}-0.40942 & -1.0508\end{bmatrix}$.
	The level sets of $\B_{v_{i}} \leq 0$ for nodes $v_i \in \{v_1, v_2, v_3\}$ and an example state sequence of the WH control system under the newly synthesized controller are depicted in Figure~\ref{fig:linear-synth-example}, omitting $\B_{v_4}$ through $\B_{v_{15}}$.
\end{casestudy}

\section{Conclusion} \label{sec:conclusion}
This article proposed a new methodology for the safety verification and controller synthesis for WH control systems, a class of discrete-time control systems that are subject to losses given by WH constraints.
Both zero and hold actuator strategies were considered.
The key idea was to leverage graph-based representations of WH constraint, and design a collection of barrier functions, called graph-based barrier function, tailored to the WH graph.
Such a framework allows to implicitly consider the allowable patterns of losses and successes encountered by the system, providing a general, non-conservative approach for checking safety specifications.
However, the resulting safety conditions may be rendered impractical due to computational intractability or larger complexity.
Therefore, we also provided alternate reformulations that are computationally more tractable, but come at the cost of higher conservatism.
A discussion on the practical computation on graph-based barrier functions was also presented.
Our numerical case studies validated the effectiveness of our results across the presented variants of graph-based barrier functions as well as actuator strategies.

\bibliographystyle{ieeetr}
\bibliography{literature.bib}

\end{document}